\documentclass[prx,twocolumn,english,superscriptaddress]{revtex4-2}

\usepackage[T1]{fontenc}
\usepackage{newtxtext}
\usepackage{newtxmath}
\usepackage{float}

\usepackage{amsmath}
\usepackage{mathtools}
\usepackage{amsfonts}
\usepackage{bm}
\usepackage{xfrac}

\usepackage{color}
\usepackage{xcolor}
\usepackage[colorlinks,
            citecolor=blue,
            linkcolor=blue,
            urlcolor=blue]{hyperref}

\usepackage{graphicx}
\usepackage[all]{hypcap} 
            
\makeatletter
\newsavebox{\@brx}
\newcommand{\llangle}[1][]{\savebox{\@brx}{\(\m@th{#1\langle}\)}%
  \mathopen{\copy\@brx\kern-0.5\wd\@brx\usebox{\@brx}}}
\newcommand{\rrangle}[1][]{\savebox{\@brx}{\(\m@th{#1\rangle}\)}%
  \mathclose{\copy\@brx\kern-0.5\wd\@brx\usebox{\@brx}}}
\makeatother

\begin{document}





\title{Non-linear magnons and exchange Hamiltonians of delafossite proximate quantum spin liquids}

\author{A. O. Scheie} 
\email{scheie@lanl.gov}
\affiliation{MPA-Q, Los Alamos National Laboratory, Los Alamos, New Mexico 87545, USA}
\affiliation{Neutron Scattering Division, Oak Ridge National Laboratory, Oak Ridge, TN 37831, USA}

\author{Y. Kamiya} 
\affiliation{School of Physics and Astronomy, Shanghai Jiao Tong University 800 Dongchuan Road, Minhang District, Shanghai 200240, China}

\author{Hao Zhang} 
\affiliation{Department of Physics and Astronomy, University of Tennessee, Knoxville, TN 37996, USA}

\author{Sangyun Lee} 
\affiliation{MPA-Q, Los Alamos National Laboratory, Los Alamos, New Mexico 87545, USA}

\author{A.J. Woods}  
\affiliation{MPA-Q, Los Alamos National Laboratory, Los Alamos, New Mexico 87545, USA}

\author{A.M. Omanakuttan}  
\affiliation{MPA-Q, Los Alamos National Laboratory, Los Alamos, New Mexico 87545, USA}

\author{M. G. Gonzalez}   
\affiliation{Sorbonne Universit\'e, CNRS, Laboratoire de Physique Th\'eorique de la Mati\`ere Condens\'ee, LPTMC, F-75005 Paris, France}

\author{B. Bernu}   
\affiliation{Sorbonne Universit\'e, CNRS, Laboratoire de Physique Th\'eorique de la Mati\`ere Condens\'ee, LPTMC, F-75005 Paris, France}

\author{J.W. Villanova}  
\affiliation{Center for Nanophase Materials Sciences, Oak Ridge National Laboratory, Oak Ridge, Tennessee 37831, USA}

\author{J. Xing} 
\address{Materials Science and Technology Division, Oak Ridge National Laboratory, Oak Ridge, Tennessee 37831, USA}

\author{Q. Huang} 
\address{Department of Physics and Astronomy, University of Tennessee, Knoxville, TN 37996, USA}

\author{Qingming Zhang} 
\address{School of Physical Science and Technology, Lanzhou University, Institute of Physics, Chinese Academy of Sciences, Lanzhou 730000, China}

\author{Jie Ma} 
\address{Department of Physics and Astronomy, Shanghai Jiao Tong University, Shanghai 200240, China}

\author{Eun Sang Choi} 
\address{National High Magnetic Field Laboratory, Florida State University, Tallahassee, Florida 32310, USA}
    
\author{D. M. Pajerowski}    
\affiliation{Neutron Scattering Division, Oak Ridge National Laboratory, Oak Ridge, TN 37831, USA}

\author{Haidong Zhou} 
\affiliation{Department of Physics and Astronomy, University of Tennessee, Knoxville, TN 37996, USA}

\author{A. S. Sefat} 
\affiliation{Materials Science and Technology Division, Oak Ridge National Laboratory, Oak Ridge, Tennessee 37831, USA}

\author{S. Okamoto}  
\affiliation{Materials Science and Technology Division, Oak Ridge National Laboratory, Oak Ridge, Tennessee 37831, USA}

\author{T. Berlijn}  
\affiliation{Center for Nanophase Materials Sciences, Oak Ridge National Laboratory, Oak Ridge, Tennessee 37831, USA}

\author{L. Messio}   
\affiliation{Sorbonne Universit\'e, CNRS, Laboratoire de Physique Th\'eorique de la Mati\`ere Condens\'ee, LPTMC, F-75005 Paris, France}
\affiliation{Institut Universitaire de France (IUF), F-75005 Paris, France}

\author{R. Movshovich} 
\affiliation{MPA-Q, Los Alamos National Laboratory, Los Alamos, New Mexico 87545, USA}

\author{C. D. Batista}   
\affiliation{Department of Physics and Astronomy, University of Tennessee, Knoxville, TN 37996, USA}
\affiliation{Shull Wollan Center, Oak Ridge National Laboratory, TN 37831. USA}

\author{D. A. Tennant}  
\affiliation{Department of Physics and Astronomy, University of Tennessee, Knoxville, TN 37996, USA}
\affiliation{Shull Wollan Center, Oak Ridge National Laboratory, TN 37831. USA}

\date{\today}

\begin{abstract}

Quantum spin liquids (QSL) are theoretical states of matter with long-range entanglement and exotic quasiparticles. However, they generally elude quantitative theory, rendering their underlying phases mysterious and hampering efforts to identify experimental QSL states. Here we study triangular lattice resonating valence bond QSL candidate materials KYbSe$_2$ and NaYbSe$_2$. We measure the magnon modes  in their 1/3 plateau phase, where quantitative theory is tractable, using inelastic neutron scattering and fit them using nonlinear spin wave theory. We also fit the KYbSe$_2$ heat capacity using high temperature series expansion. Both KYbSe$_2$ fits yield the same magnetic Hamiltonian to within uncertainty, confirming previous estimates and showing the Heisenberg $J_2/J_1$ to be an accurate model for these materials. 
Most importantly, comparing KYbSe$_2$ and NaYbSe$_2$ shows that smaller $A$-site Na$^+$ ion has a larger $J_2/J_1$ ratio. However, hydrostatic pressure applied to KYbSe$_2$ increases the ordering temperature (a result consistent with density functional theory calculations), indicating  that pressure decreases $J_2/J_1$. These results show how periodic table and hydrostatic pressure can tune the $A$YbSe$_2$ materials in a controlled way. 
    \end{abstract}

\maketitle

\section{Introduction} 

Triangular lattice quantum magnets have been of intense interest since Anderson's 1973 prediction of a resonating valence bond (RVB) quantum spin liquid (QSL) \cite{Anderson1973}, 
but despite years of searching, no unambiguous triangular QSL materials exist \cite{broholm2019quantum,Syzranov2022}. In this state, antiferromagnetic interacting spins on a two-dimensional triangular lattice produce a long-range entangled quantum spin liquid rather than conventional long range order \cite{Balents2010review}. 
This state has intriguing theoretical properties \cite{Zhou_2017_Quantum}, and quantum spin liquids holds great potential for quantum electronic technology \cite{Nayak_2008,tokura2017emergent}.
Although Anderson's original QSL proposal was for the nearest neighbor Heisenberg antiferromagnet, subsequent studies showed this model actually orders into a $120^{\circ}$ phase at the lowest temperatures \cite{Capriotti_1999,White_2007}. Instead, the QSL state requires a small second neighbor $J_2$ exchange between $\approx 6$\% and $\approx 16$\% of the nearest neighbor exchange to stabilize \cite{PhysRevB.92.041105,PhysRevB.92.140403,PhysRevB.93.144411,PhysRevB.94.121111,PhysRevB.95.035141,PhysRevB.96.075116,PhysRevLett.123.207203}. (A similar role is played by nearest neighbor anisotropic exchange~\cite{Zhu_2018}.)
Although it is unclear whether this $J_2$-stabilized phase is Anderson's RVB or a different type of QSL \cite{PhysRevLett.123.207203,PhysRevB.100.241111}, it is clear that the QSL state should exist---
if the Hamiltonian parameters can be tuned appropriately. 

Many materials have been proposed as 2D triangular lattice antiferromagnets, including ${\mathrm{Ba}}_{3}{\mathrm{CoSb}}_{2}{\mathrm{O}}_{9}$ \cite{Ito2017,Macdougal_2020,Ma_2016}, ${\mathrm{YbMgGaO}}_{4}$ \cite{Li_2015_YMGO,Shen2016,Paddison2017,Xu_2016_YMGO,Zhu_2017_YMGO}, and organic salts \cite{Itou2008,Yamashita2008,Riedl2019}. However, none of these have been shown to have an RVB ground state. 
A promising new class of materials is the Yb$^{3+}$ delafossites, which have magnetic Yb$^{3+}$ in a crystallographically perfect 2D triangular lattice \cite{Ding_2019_NYO,Baenitz_2018,Bordelon2019,sarkar2019quantum,Xing2021_KYS,Dai_2021}. 
However, the crucial test in evaluating candidate RVB materials is  whether their magnetic exchange Hamiltonians are indeed within a theoretical QSL phase.

\begin{figure*}
	\centering\includegraphics[width=\textwidth]{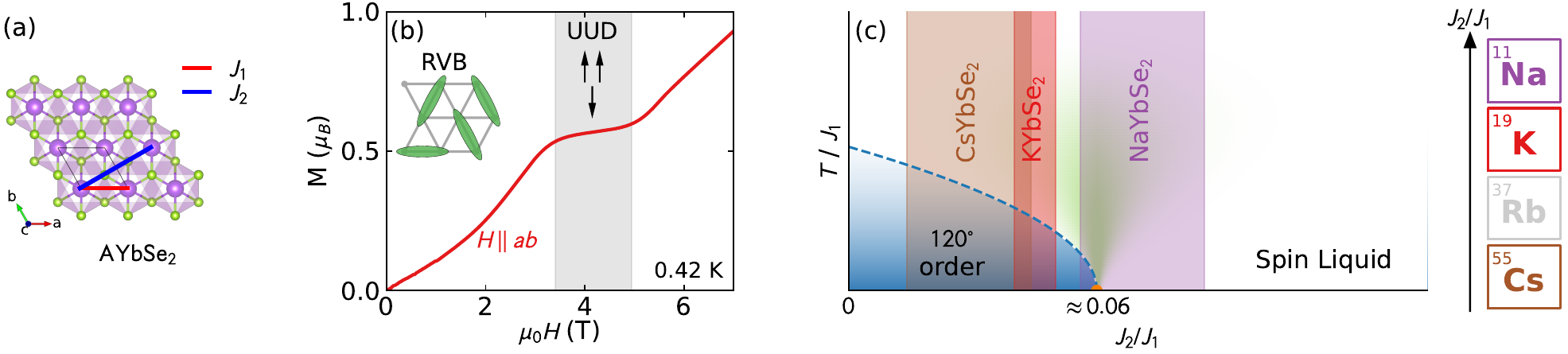}
	\caption{AYbSe$_2$ crystal structure and phase diagram. (a) The triangular lattice plane with the $J_1$ and $J_2$ exchanges between magnetic Yb sites. (b) The 0.42~K in-plane KYbSe$_2$ magnetization from Ref. \cite{Xing2021_KYS}, showing a 1/3 magnetization plateau phase at 4~T. (c) The $J_2/J_1$ phase diagram, showing the refined AYbSe$_2$ $J_2/J_1$ for A$=$Cs, K, and Na. As the A site moves up the periodic table, the $J_2/J_1$ increases and the Hamiltonian moves closer to the spin liquid state.}
	\label{flo:Schematic}
\end{figure*}

Part of the difficulty in experimentally studying an RVB liquid is that its lacks sharp spectral features: its elementary quasiparticles are $S=1/2$ spinons, which are created in pairs and thus produce a diffuse continuum \cite{Kivelson_1987,Anderson_1987}. And unfortunately, diffuse continua are also produced by disordered or glassy states and are thus ambiguous \cite{Zhu_2017_YMGO,Kimchi_2018}. 
However, an important feature of the quantum $S=1/2$ triangular lattice antiferromagnet is that an applied magnetic field produces a 1/3-magnetization plateau phase corresponding to up-up-down long range magnetic order \cite{Nishimori_1986,Chubukov_1991}. This is an inherently quantum mechanical effect, whereby quantum fluctuations select a collinear spin ordering \cite{Alicea_2009}. Crucially, this produces well-defined  magnon modes within the plateau phase,  whose gapped dispersion can be calculated  via nonlinear spin wave theory \cite{Alicea_2009,Kamiya18}. This  observation allows us to extract the Hamiltonian parameters without the need for reaching the fully polarized magnetic state (the saturation field is prohibitively large for most trialgular QSL candidate materials \cite{Coldea_2002,Kamiya18}). The problem thus becomes tractable by studying the magnons in the plateau phase. 

KYbSe$_2$ \cite{Xing2021_KYS} and NaYbSe$_2$ \cite{Liu_2018_Chalcogenides} are recently discovered 2D triangular QSL candidate materials.
Like the ideal 2D triangular  Heisenberg model \cite{Nishimori_1986,Chubukov_1991}, KYbSe$_2$ and  NaYbSe$_2$ have a 1/3 saturation magnetization plateaus at in-plane magnetic fields of 4~T [Fig. \ref{flo:Schematic}(b)] and 4.7~T \cite{Ranjith2019_2} respectively, which also accompany a reentrant magnetic ordered phase \cite{Xing2021_KYS}. Presumably, this ordered phase is the up-up-down phase of the quantum $S=1/2$ triangular lattice antiferromagnet. This finite field ordered phase is itself evidence of a highly quantum ground state, and also provides an opportunity to observe coherent spin wave modes at modest magnetic field, and fit them to a non-linear spin wave theory (NLSWT).

KYbSe$_2$ was recently studied using neutron scattering, and fits to paramagnetic diffuse scattering show it to be a proximate QSL: it orders magnetically at 290~mK, but its fitted $J_2/J_1$ ratio is close enough to the QSL phase to exhibit some of its exotic behaviors (see Fig. \ref{flo:Schematic}), including diffuse zero-field neutron spectra from bound spinons \cite{scheie2021_KYS}. In NaYbSe$_2$, meanwhile, no magnetic order has been reported \cite{Ranjith2019_2,sarkar2019quantum,Zhang_2022_NYS} and its neutron spectra is also diffuse \cite{Dai_2021}. 
However, the previous KYbSe$_2$ fit relied upon classical models compared to paramagnetic scattering, and the NaYbSe$_2$ study did not refine the $J_2/J_1$ value. A robust Hamiltonian fit is required for both materials. 
Here we provide such analysis using inelastic magnetic scattering and heat capacity, showing  KYbSe$_2$ and NaYbSe$_2$ to be remarkably good realizations of the $J_2/J_1$ triangular lattice Heisenberg antiferromagnet. Furthermore, the comparison between K and Na on the $A$-site in $A$YbSe$_2$ shows that replacing K by the smaller Na ion gives a larger $J_2/J_1$ ratio. We hypothesized that hydrostatic pressure would similarly push KYbSe$_2$ towards a QSL, but measurements indicate that it increases the ordering temperature rather than decreasing it. 
Nonetheless, these results show how one can rationally tune triangular lattice materials to realize a QSL phase.

\section{Experiments and Results}

\subsection{Neutron Scattering}

\begin{figure*}
	\centering\includegraphics[width=0.96\textwidth]{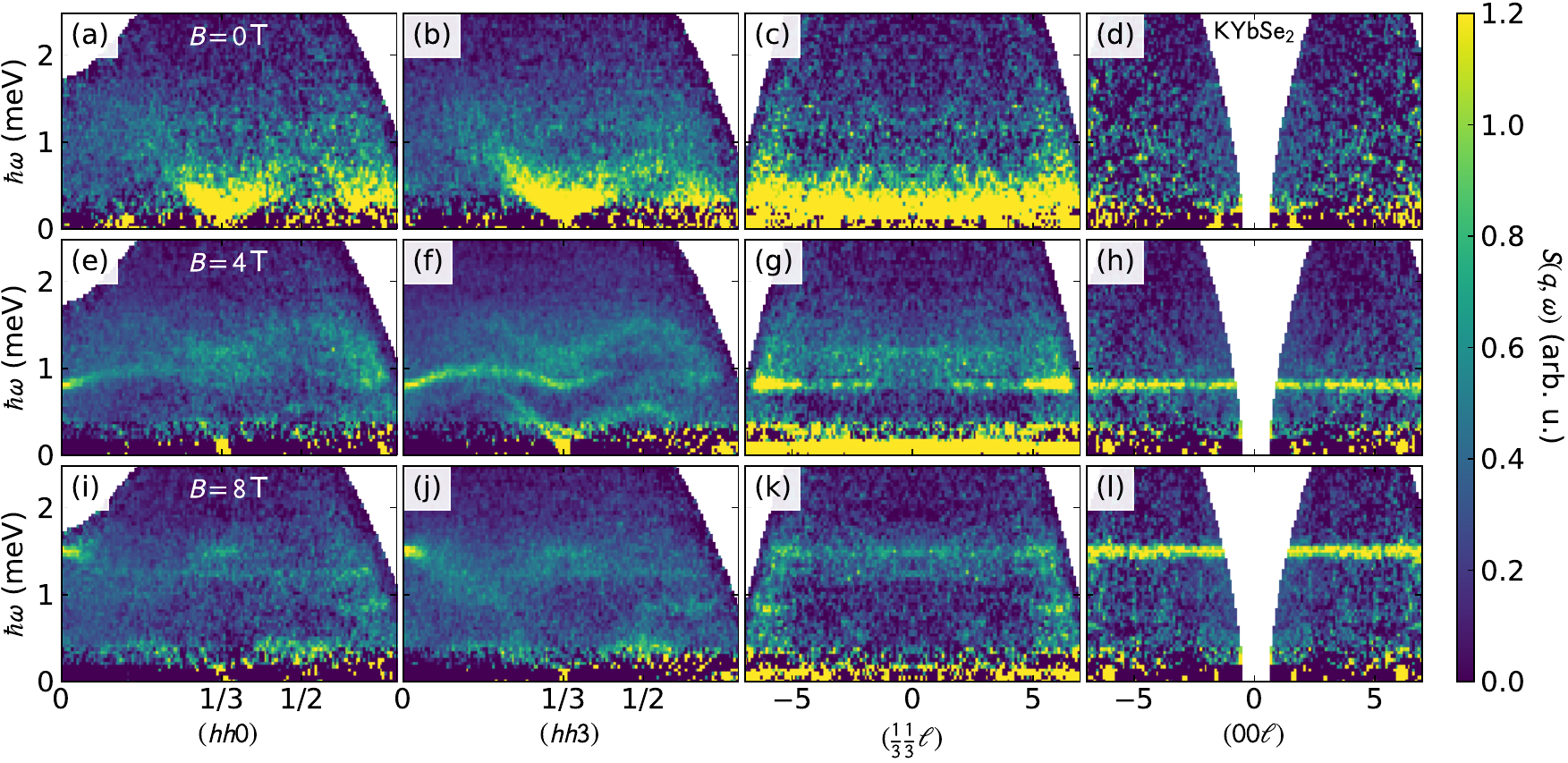}
	\caption{KYbSe$_2$ field dependent scattering. The top row shows the zero field neutron spectrum, showing a diffuse continuum with a lower bound coming to zero energy at $hh=(1/3,1/3)$ (in reciprocal lattice units). The middle row shows the same slices for $B=4$~T and the bottom row for $B=8$~T. The left two columns show cuts along $hh$ (the far left integrated over $-1.5 < \ell < 1.5$ r.l.u., the center-left column integrated over $1.5 < \ell < 4.5$), and the right two columns show cuts along $\ell$. For $B=0$~T and $B=8$~T, there is no appreciable $\ell$ dependence to the scattering, but at $B=4$~T there is a modulated intensity to some of the features as shown most clearly in panel (g).} 
	\label{flo:NeutronScattering}
\end{figure*}

\begin{figure}
	\centering\includegraphics[width=0.48\textwidth]{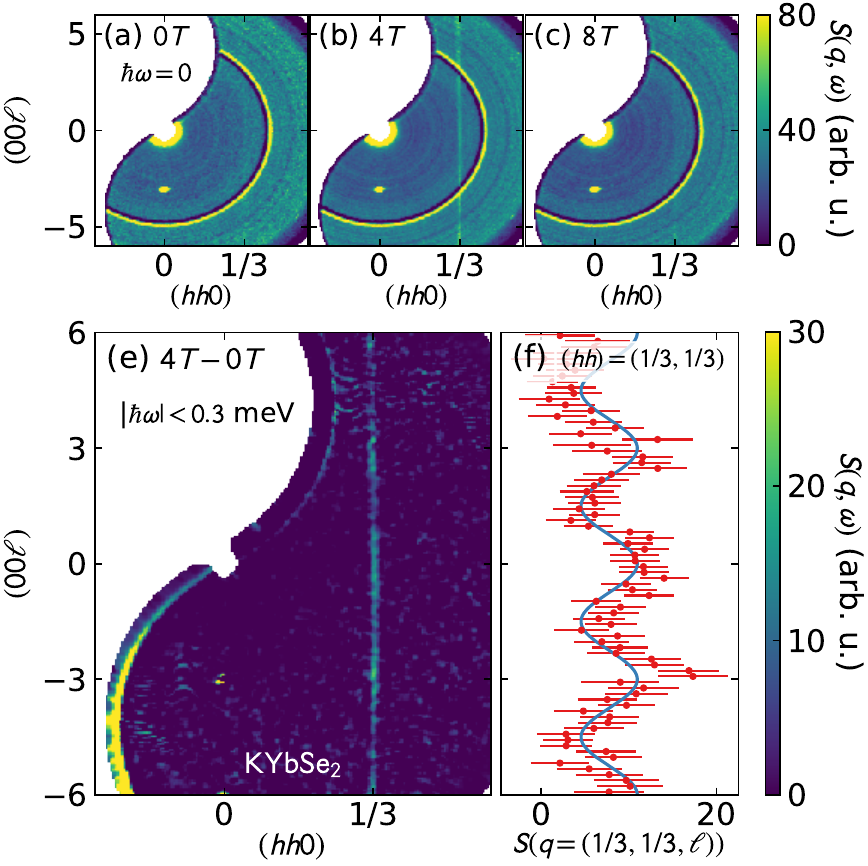}
	\caption{KYbSe$_2$ elastic scattering along $\ell$. Panels (a) - (c) show the elastic scattering ($\hbar \omega < 0.3$~meV) at 0~T, 4~T, and 8~T. Any elastic scattering extremely weak at 0~T and 8~T, but at 4~T a vertical streak appears at $(1/3,1/3,\ell)$, indicating static magnetism. Panel (e) shows the 4~T with 0~T subtracted as a background, revealing a sinusoidal modulation in elastic intensity, which is plotted in panel (f) with the blue line as a guide to the eye. This evidences static magnetism in the plateau phase which is only weakly correlated between planes.}
	\label{flo:Ldependence}
\end{figure}

\begin{figure*}
	\centering\includegraphics[width=0.92\textwidth]{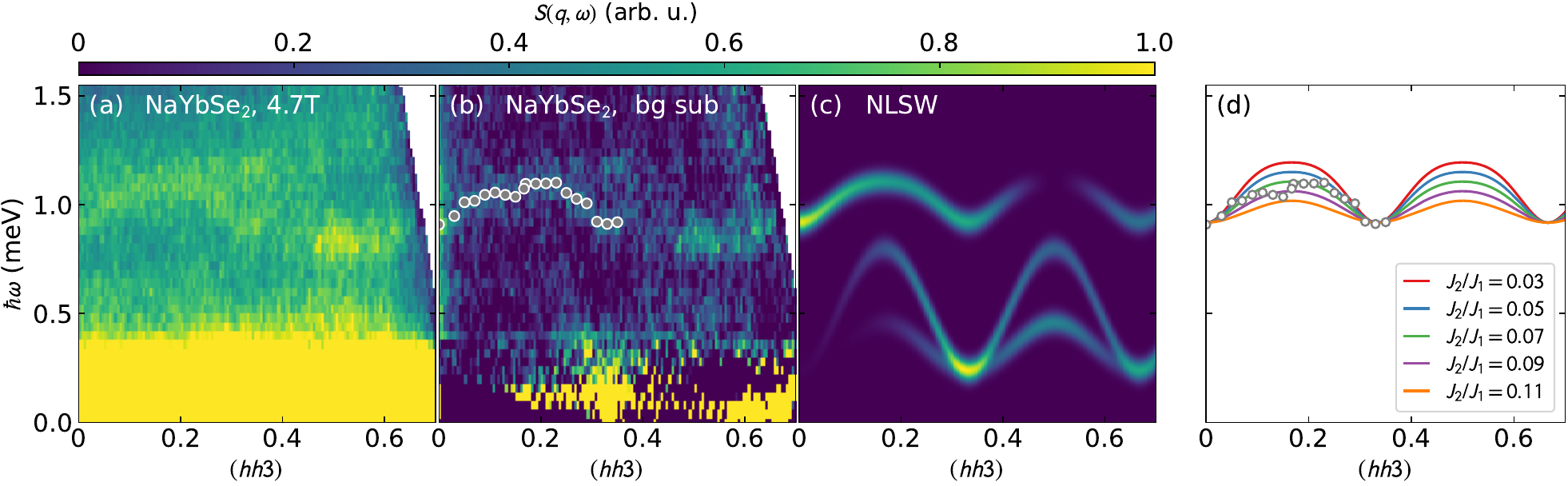}
	\caption{Nonlinear spin wave fit to NaYbSe$_2$ scattering at 4.7~T. Panel {\bf a} shows the experimental scattering along ($hh3$), and panel {\bf b} shows the same data with the background subtracted. Panel {\bf c} shows the best calculated neutron spectrum from nonlinear spin wave theory (NLSWT). Panel {\bf d} shows the mode centers determined using constant $Q$ cuts fitted to a Gaussian profile in energy, compared with with the NLSWT dispersion curves used to fit the data. Despite the low statistics, the bandwidth of the upper mode constrains $J_2/J_1$ to within $\pm0.015$.}
	\label{flo:NYS_NLSWfit}
\end{figure*}

We measured the KYbSe$_2$ and NaYbSe$_2$ scattering in the $hhl$ scattering plane using the CNCS spectrometer~\cite{CNCS} at Oak Ridge National Laboratory's Spallation Neutron Source~\cite{mason2006spallation}.  The data for KYbSe$_2$ are shown in Figs. \ref{flo:NeutronScattering} and \ref{flo:Ldependence}, and the plateau field data for NaYbSe$_2$ are shown in Fig. \ref{flo:NYS_NLSWfit} (the low field NaYbSe$_2$ scattering will be published in a separate paper).




The zero field  KYbSe$_2$ neutron scattering in Fig. \ref{flo:NeutronScattering} shows a diffuse continuum with a sharp lower bound becoming gapless at $(1/3,1/3,0)$ as reported previously \cite{scheie2021_KYS}. When the magnetic field is set to within the plateau field (4~T), well-defined spin wave modes appear, similar to what was reported in CsYbSe$_2$ \cite{xie2021field,xie2022complete}. When the magnetic field is increased further beyond the plateau field, the well-defined modes disappear, replaced again by broad, diffuse features---although there are sharply defined features at $hh= (0,0)$ and $(1/3,1/3)$ at 8~T as shown in Fig. \ref{flo:NeutronScattering}(i)-(l).

At all fields, there is no detectable dispersion in the $\ell$ direction (right columns of Fig. \ref{flo:NeutronScattering}), signaling highly two-dimensional magnetism and very weak interplane exchange. However, at 4~T there is noticeable intensity modulation in the modes, as shown in Fig. \ref{flo:NeutronScattering}(g) and Fig. \ref{flo:Ldependence}. At 4~T, a streak of elastic scattering appears at $(1/3,1/3,\ell)$, in accord with in-plane magnetic order of the plateau phase. If we isolate the magnetic scattering by subtracting the 0~T data (where the magnetic order is extremely weak and barely any elastic magnetic signal is visible) from the 4~T data as in Fig. \ref{flo:Ldependence}(e), we see that the 4~T magnetic signal has a sinusoidal intensity modulation with $\ell$, as shown in Fig. \ref{flo:Ldependence}(f). The intensity peaks at $\ell=0$ and $\ell=\pm 3$, the same periodicity as the triangular planes, signaling short-ranged ferromagnetic correlations between the triangular lattice planes with a propagation vector $Q=(1/3,1/3,0)$.
This is markedly different from the elastic magnetic scattering at 4~T measured in CsYbSe$_2$, which had maximal intensity at $\ell=\pm 1$ and a propagation vector $Q=(1/3,1/3,1)$ \cite{xie2021field}. Furthermore, in CsYbSe$_2$ the correlations extend to at least three triangular lattice planes, but in KYbSe$_2$ the magnetic correlations appear to extend only to the neighboring planes. Thus KYbSe$_2$ is more two-dimensional than CsYbSe$_2$.

\begin{figure*}
	\centering\includegraphics[width=0.96\textwidth]{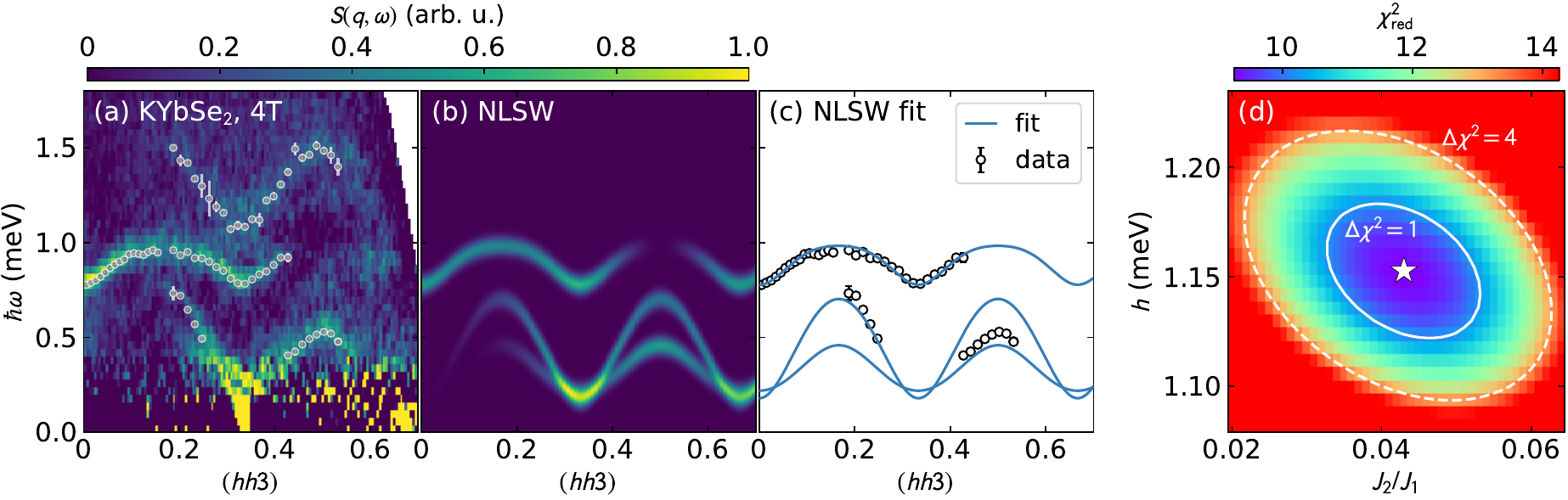}
	\caption{Nonlinear spin wave fit to KYbSe$_2$ scattering at 4~T. The left panel shows the experimental scattering integrated along ($hh3$), with the grey data points indicating the mode centers determined using constant $Q$ cuts fitted to a Gaussian profile in energy. Panel (b) shows the best calculated neutron spectrum from nonlinear spin wave theory (NLSWT). Note that the highest energy mode is not captured by the single-magnon scattering calculation. Panel (c) shows the fitted points from panel (a) with the NLSWT dispersion curves used to fit the data. Panel (d) shows the one and two standard deviation $\chi^2$ contours of $J_2/J_1$ and magnetic field $h$.}
	\label{flo:NLSWfit}
\end{figure*}

Turning to the inelastic scattering, the well defined modes at 4~T provide an opportunity to fit the magnetic exchange Hamiltonian using nonlinear spin wave theory (NLSWT) \cite{Alicea_2009,Kamiya18}. This provides an independent test of whether the $J_1$-$J_2$ Hamiltonian is appropriate to KYbSe$_2$ and NaYbSe$_2$, and what the $J_2/J_1$ ratio is.

We assume a $J_1$-$J_2$ Hamiltonian in a magnetic field
\begin{equation}
    \hat{\mathcal{H}} = J_1\sum_{\langle i,j\rangle}\hat{\boldsymbol{S}}_i\cdot\hat{\boldsymbol{S}}_j+J_2\sum_{\langle\langle i,j\rangle\rangle}\hat{\boldsymbol{S}}_i\cdot\hat{\boldsymbol{S}}_j  - h \sum_{i}  \hat{S}^x_i
    \label{eq:hamiltonian}
\end{equation}
 where   $h = g_{xx} \mu_B B$, B is the magnitude of the applied magnetic field along the $x$-direction and $g_{\alpha \beta}$ is the $g$-tensor.


By following the procedure described in Ref.~\cite{Kamiya18}, we fit the NLSWT magnon dispersion to the  mode energies as a function of wavevector extracted from Gaussian profiles at constant $Q$ slices [see Fig. \ref{flo:NLSWfit}(a) and \ref{flo:NYS_NLSWfit}(b)]. 
For KYbSe$_2$, we treated the three Hamiltonian parameters  $J_1$, $J_2$, and   $h$  as fitted constants (magnetic field $h$ is a fitted parameter because 
$g_{xx}$ has large uncertainty for KYbSe$_2$ \cite{scheie2021_KYS,scheie2022quantifying}) and fit the NLSWT mode energies to the experimentally measured dispersions as shown in Fig. \ref{flo:NLSWfit}(c). This gives a very good reproduction of the lower energy modes, as shown in Fig. \ref{flo:NLSWfit}(b). To define uncertainty, we calculated a contour $\Delta \chi^2_{\rm red} = 1$ above the reduced $\chi^2$ minimum \cite{NumericalRecipes}. One standard deviation and two standard deviation contours are plotted in Fig. \ref{flo:NLSWfit}(d). Using a one standard deviation uncertainty, the best fit values are $J_1 = (0.456 \pm 0.013)$~meV, 
$J_2/J_1 = 0.043 \pm 0.010$, and $h/J_1 = 1.73 \pm 0.05$.
This $J_2/J_1$ ratio agrees to within uncertainty with the $J_2/J_1 = 0.047 \pm 0.007$ from the Onsager reaction field fits in Ref. \cite{scheie2021_KYS}.  We note that the exact saturation field for the proposed model is $h_{\rm sat} /J_1= 9/2$, implying that $h/h_{\rm sat}=0.38 \pm 0.01$. 

For NaYbSe$_2$, because of the larger background from the copper sample mount, only one magnon mode is clearly visible in the data. Nevertheless, this mode is clearly at higher energy than the same mode in KYbSe$_2$ and thus contains meaningful information. To constrain this fit, we used the measured saturation field to constrain $h$ (see Appendix \ref{app:Experiments}). This allows for $J_1$ (determined by the height of the mode at $hh=0$) and $J_2/J_1$ (determined by the bandwidth of the mode). Using the same method of $\chi^2$ contours to define uncertainty, we find a fitted $J_1 = (0.551 \pm 0.010)$~meV and $J_2/J_1 = 0.071 \pm 0.015$. Although the error bars are large, both $J_1$ and $J_2/J_1$ are larger in NaYbSe$_2$ than KYbSe$_2$. 

 Single-magnon NLSWT calculations did not capture the KYbSe$_2$  dispersive feature above 1~meV in Fig. \ref{flo:NLSWfit}(a). However, two-magnon continuum calculations (shown in Appendix \ref{app:NonLinearSpinWaves}) do capture this feature. The sharpness of this mode in an otherwise diffuse continuum arises from the two-dimensional nature of the system, such that the density of two-magnon states has Van Hove singularities at energies that depend on their total momentum. These  singularities are broadened by the finite experimental resolution, leading  to a dispersive peak in the  continuum scattering arising from the longitudinal spin structure factor, as was seen in YbCl$_3$~\cite{sala2021van}.  

\subsection{Heat Capacity}

\begin{figure}
	\centering\includegraphics[width=0.48\textwidth]{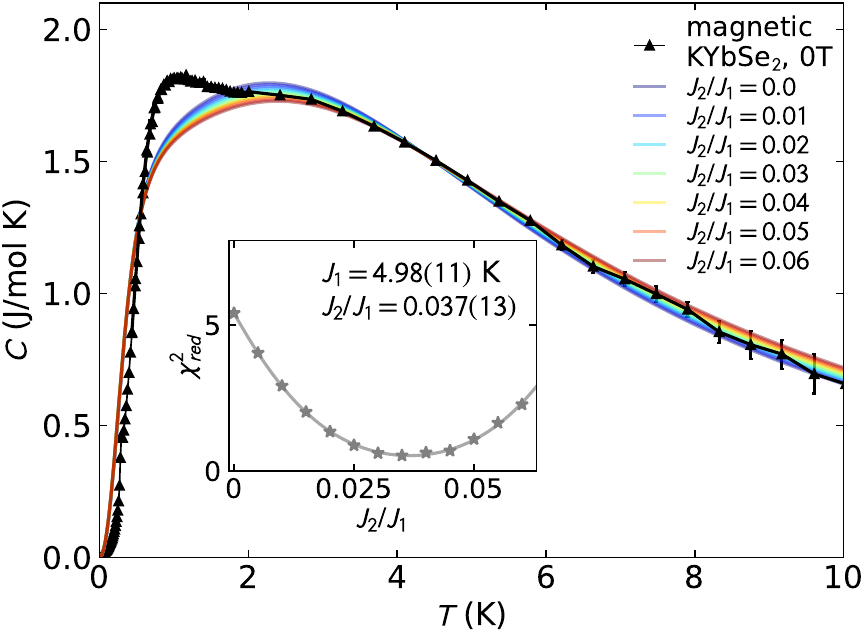}
	\caption{KYbSe$_2$ zero-field heat capacity (black) compared to theoretical calculations of the heat capacity from Ref. \cite{gonzalez2021ground}. 
	For the theory, the only adjustable parameter is the $T$ axis scaling (corresponding to $J_1$). The inset shows reduced $\chi^2$ as a function of $J_2/J_1$, showing a minimum at $J_2/J_1=0.040(12)$.}
	\label{flo:HeatCapacity}
\end{figure}

As another independent test of our $J_2/J_1$ model, we also compared the KYbSe$_2$ heat capacity to theoretical calculations. Using the zero field heat capacity reported in Refs.~\cite{Xing2021_KYS,scheie2021_KYS}, subtracted by KLuSe$_2$ heat capacity to isolate the magnetic heat capacity (see Appendix \ref{app:Experiments} for details), we compared the data to the high temperature series expansion interpolation calculations \cite{Bernu_2001,Schmidt_2017,Bernu_2020,gonzalez2021ground}, which give accurate quantum calculations of the high temperature heat capacity of the 2D triangular lattice as a function of $J_2/J_1$.  The series were calculated out to the 14th order (one order more than Ref. \cite{gonzalez2021ground}). These data are shown in Fig. \ref{flo:HeatCapacity}.

For temperatures greater than 2~K, the theory and experiment match very closely, with subtle differences depending on the precise $J_2/J_1$ value. To quantitatively compare theory and experiment, we define a $\chi^2_{red}$ for $2\>{\rm K} < T < 8\>{\rm K}$. (8~K is where the lattice heat capacity becomes much larger than the magnetic contribution, and the subtracted values are more questionable.) For each theoretical $J_2/J_1$ value, we fit $J_1$ to minimize $\chi^2$ and thus obtain $\chi^2_{\rm red}$ as a function of $J_2/J_1$, shown as an inset to Fig. \ref{flo:HeatCapacity}.
This fit yields best fit values of $J_1=(0.439 \pm 0.010)$~meV and $J_2 /J_1 = 0.037 \pm 0.013$. Both values agree to within uncertainty with the results from NLSWT fits. 

It should be noted that the high temperature expansion does not capture either the maximum in heat capacity at 1~K or the ordering transition at 290~mK. The lack of ordering transition is not surprising as the theory is based off a perfectly isotropic 2D lattice, which only orders at $T=0$ \cite{MerminWagner}. The 1~K ``bump'', also seen in NaYbO$_2$ \cite{Bordelon2019} and NaYbSe$_2$ \cite{Ranjith2019_2}, is more challenging to explain. Finite system tensor network calculations \cite{Chen_2019_twotemperature} showed a two-hump heat capacity for the 2D triangular lattice antiferromagnet
---but it is not clear whether this hump evolves to the sharp feature 290~mK or the 1~K hump in KYbSe$_2$. Be that as it may, the failure of the high-temperature expansion in capturing this peak  suggests that it is either induced by an increase of the magnetic in-plane correlation length beyond several lattice spaces, $\xi \gtrsim 10 a$, or by the onset of  three-dimensional correlations in KYbSe$_2$, which ultimately leads to long range magnetic order. 
In any case, the series expansion theory gives an excellent description of the $> 2$~K heat capacity, and allows us to extract both $J_1$ and $J_2$ from the data.

To test whether hydrostatic pressure can tune KYbSe$_2$ further towards a QSL phase, we measured AC calorimetry under applied hydrostatic pressure, plotted in Fig. \ref{flo:HeatCapacityPressure} (see Appendix \ref{app:Experiments} for experimental details). 
We found that the ordering transition was clearly visible, but applied pressure shifted $T_N$ to higher temperatures. This signals that the system is getting further from a QSL, as the 120$^{\circ}$ order becomes more stable. This could be from a decrease in $J_2/J_1$ or perhaps an increase in the interlayer exchange strength which stabilizes the long range magnetic order---but in either case, hydrostatic pressure tunes KYbSe$_2$ \textit{away} from a QSL. 

\begin{figure}
	\centering\includegraphics[width=0.5\textwidth]{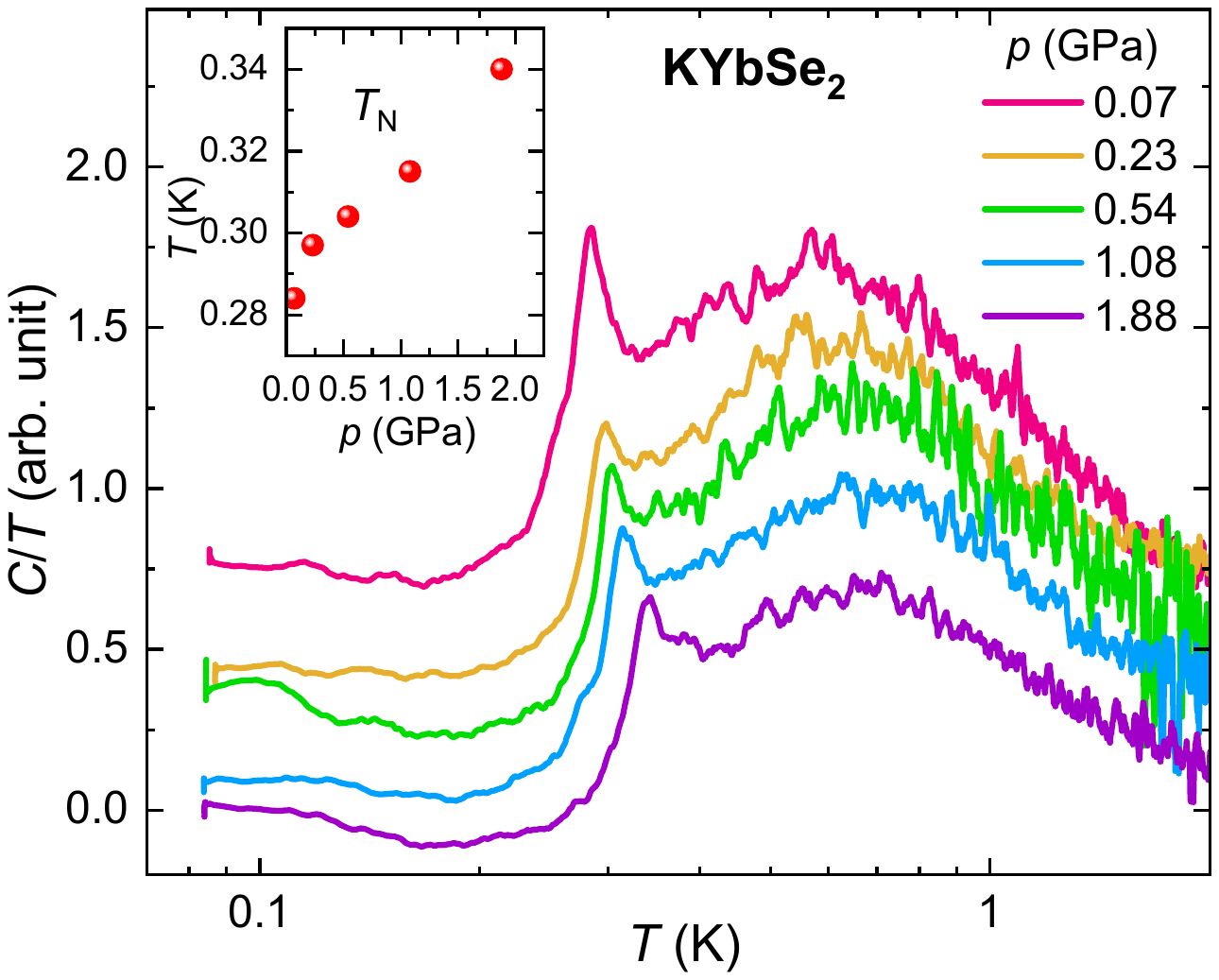}
	\caption{Pressure dependent KYbSe$_2$ specific heat (in arbitrary units), with increasing pressures offset $-C/T$ for clarity. The inset shows the temperature of the ordering transition peak center. As pressure increases, the transition noticeably increases, signaling a move away from the QSL phase.}
	\label{flo:HeatCapacityPressure}
\end{figure}

\subsection{Density Functional Theory}

To better understand these experimental results, we use density functional theory (DFT), Wannier functions, and strong coupling perturbation theory to derive the exchange constants at various pressures for KYbSe$_2$. The results are presented in Table~\ref{ta:expressure}, and calculation details are in Appendix \ref{app:DFT}. The nearest neighbor exchange $J_1$ is in reasonable agreement with the experimental weighted mean fit, and increasing pressure tends to increase $J_1$. Although the theoretical second nearest neighbor exchange constants are an order of magnitude smaller compared to the experimental observations, they do display the correct trend decreasing as a function of pressure. Likewise the ratio $|J_2/J_1|$ decreases with increasing pressure. In addition, Table~\ref{ta:expressure} shows the pressure dependence of the third nearest in-plane neighbor exchange ($J_3$), the nearest out-of-plane neighbor exchange ($J_{1}^{\rm out}$), the nearest neighbor Kitaev interaction ($K_1$), and the nearest neighbor bond-dependent anisotropic exchange parameters ($\Gamma_1,\Gamma^{\prime}_1$). We refer to Ref.~\cite{Villanova2023} for further details. 
This suggests that the experimental observation that hydrostatic pressure tunes KYbSe$_2$ away from the QSL phase is really because hydrostatic pressure decreases $|J_2/J_1|$. 


\begin{table}[htb!]
\begin{center}
\begin{ruledtabular}
\begin{tabular}{c|c|c|c}
KYbSe$_2$ & 0 GPa  & 2 GPa  & 5 GPa  \\
\hline
$J_1$ (meV) & 0.428 & 0.480 & 0.494  \\
$J_2$ ($\mu$eV) & 2.183 & 1.967 & 0.583  \\
$J_3$ ($\mu$eV) & 5.036 & 4.412 & 1.237 \\ 
$J^{out}_{1}$ (meV) & 0.016 & 0.016 & 0.016 \\
\hline
$J_2/J_1$ & 0.005 & 0.004 & 0.001 \\ 
$J^{out}_{1}/J_1$ & 0.038 & 0.033 & 0.033 \\ 
\hline
$K_1$ ($\mu$eV) & 14.457 & 15.692 & 18.330   \\
$\Gamma_1$ ($\mu$eV) & $-11.213$ & $-10.851$ & $-13.730$  \\
$\Gamma^{\prime}_1$ ($\mu$eV) & $-1.831$ & $-1.577$ & $-4.526$   \\
\end{tabular}
\end{ruledtabular}
\end{center}
\caption{First principles calculations of the KYbSe$_2$ exchange constants as a function of pressure.}
\label{ta:expressure}
\end{table}

\section{Discussion}

At this point, we have three independent fits of the KYbSe$_2$ magnetic exchange Hamiltonian, which are summarized in Table \ref{tab:BestFitParameters}. Each fit used a different experiment and a different theoretical technique, so it is remarkable that all three agree to within uncertainty. The two theoretical fits performed in this study (NLSWT and high temperature heat capacity series expansion) include quantum effects, and are appropriate for the highly quantum KYbSe$_2$. Even so, the Onsager reaction field theory fit agrees to within uncertainty with these two other methods even though it neglects quantum effects. This is evidence that fitting paramagnetic scattering with classical methods \cite{Paddison_2020} works even for highly quantum systems. Also, this is a strong confirmation of the results and methodology in Ref. \cite{scheie2021_KYS}. 
All this means that KYbSe$_2$ is an excellent example of a spin-$1/2$ $J_2/J_1$ Heisenberg magnet on a triangular lattice. The use of nonlinear spin-wave theory at an intermediate---rather than fully saturated---field provides an alternative method to that in Ref. \cite{Coldea_2002} 
for triangular materials where the saturation field cannot be reached. 

\begin{table}
	\caption{Best fit exchange Hamiltonian value for KYbSe$_2$ from three different experiments and three different theoretical techniques. Note that the Onsager reaction field theory from Ref. \cite{scheie2021_KYS} cannot estimate $J_1$ as it is classical. The last row gives the weighted mean.}
	\centering
	\begin{ruledtabular}
		\begin{tabular}{c|cc}
			Theoretical technique & $J_1$~(meV) & $J_2/J_1$ \\
			\hline
			Onsager reaction field & \textit{NA} & $0.047 \pm 0.007$ \\
			Nonlinear spin waves & $0.456 \pm 0.013$ & $0.043 \pm 0.010$ \\
			Heat capacity & $0.429 \pm 0.010$ & $0.037 \pm 0.013$ \\
			\hline
			Weighted mean: & $0.438 \pm 0.008$ & $0.044 \pm 0.005$
	\end{tabular}\end{ruledtabular}
	\label{tab:BestFitParameters}
\end{table}


 Perhaps most important is the comparison between different $A$YbSe$_2$ compounds, as is visually shown in Fig. \ref{flo:Schematic}. CsYbSe$_2$ has a fitted $J_2/J_1 = 0.029 \pm 0.015$ \cite{xie2022complete}, and we show here a fitted NaYbSe$_2$ of $J_2/J_1 = 0.071 \pm 0.015$. This means that as the $A$-site ionic radius decreases (i.e., moving $A$ up the periodic table), $J_2/J_1$ increases (Fig. \ref{flo:Schematic}). This correlation gives the ability to control the material Hamiltonian by simply changing elements. 
 The AC calorimetry and DFT results show that that hydrostatic pressure also controls $J_2/J_1$, but at least in KYbSe$_2$ the shift is to smaller $J_2$ and thus further from the QSL phase. 
 
Does this mean that NaYbSe$_2$ is a QSL? Unfortunately, a fitted $J_2/J_1$ mostly within a theoretical QSL phase is insufficient evidence for such a claim. Firstly, the precise boundary between 120$^{\circ}$ and QSL order is only approximately known theoretically, and the true phase boundary is somewhere close to $J_2/J_1 \approx 0.06$. Secondly, the phase diagram in Fig. \ref{flo:Schematic}(c) neglects three-dimensional and anisotropic effects which may induce long range order in what would otherwise be a 2D QSL, or else modify the fitted $J_2/J_1$ value. (And importantly, the comparison between  CsYbSe$_2$ and KYbSe$_2$ $\ell$-dependent scattering shows the interplane exchange varies with the $A$-site element. Furthermore, the absence of a  NaYbSe$_2$ magnetization plateau with an out-of-plate field could indicate planar anisotropy in the Na compound \cite{Ranjith2019_2}.)  Thirdly, published heat capacity of NaYbSe$_2$ shows a Yb$^{3+}$ nuclear schottky anomaly at the lowest temperatures \cite{Zhang_2021_Effective,Ranjith2019_2} (though not in Ref. \cite{Dai_2021}, whose samples had 3\% Yb mixing disorder). A schottky anomaly may indicate static Yb$^{3+}$ magnetism \cite{scheie2021_KYS}, and thus weak 120$^{\circ}$ order rather than a QSL phase in the cleanest NaYbSe$_2$ samples. 
Despite such caveats, this study's confirmation that KYbSe$_2$ and NaYbSe$_2$ are very close to a 
 quantum spin liquid state provides strong impetus to further study this class of compounds as a rational route to quantum spin liquids.
  NaYbSe$_2$ may be the long-sought QSL, but even if it is not, further substitution of $A$-site and Se-site elements may allow even larger $J_2/J_1$ values to be achieved, potentially creating a genuine QSL ground state. Furthermore, if one could ``overshoot'' the QSL phase using chemical substitution, our results show that hydrostatic pressure could be used to tune  $J_2/J_1$ to a desirable value. 
 
\section{Conclusion}

We have used quantum simulation techniques to fit the KYbSe$_2$ and  NaYbSe$_2$ magnetic exchange Hamiltonians to experimental data: nonlinear spin waves to finite-field scattering in a plateau phase, and high temperature expansions to zero field heat capacity (for KYbSe$_2$). Both give extremely good agreement with experiment (down to an energy scale of 2~K), showing that the Heisenberg $J_2/J_1$ model is an excellent minimal model for KYbSe$_2$ and NaYbSe$_2$.  With three independent experimental  measurements all giving the same result for  KYbSe$_2$, this definitively verifies the model proposed in Ref. \cite{scheie2021_KYS}. 
Furthermore, both materials have a $J_2/J_1$ ratio extremely close to the 2D triangular lattice  spin liquid state at $J_2/J_1 \gtrapprox 0.06$. What is more, the replacement of K$^+$ by the smaller ion Na$^+$ increases $J_2/J_1$ and pushes the system closer to a QSL phase. Comparison to CsYbSe$_2$ shows that this trend holds across the left column of the periodic table. 
Application of hydrostatic pressure also tunes the $J_2/J_1$ ratio, but towards smaller  $J_2$. If a compound were found in the QSL phase, pressure could be used to tune it across the quantum phase transition to magnetic order. 
This not only gives very strong evidence of the appropriateness of the Heisenberg $J_2/J_1$ model and proximate QSL behavior for $A$YbSe$_2$ materials, it gives a route for engineering the magnetic Hamiltonian directly toward a QSL phase.


\begin{acknowledgments}

The work by A. Scheie, H. Zhang, S. Lee, A.J. Woods, A.M. Omanakuttan, J.W. Villanova, S. Okamoto, T. Berlijn, R. Movshovich, C.D. Batista, and D.A. Tennant is supported by the Quantum Science Center (QSC), a National Quantum Information Science Research Center of the U.S. Department of Energy (DOE). 
The neutron scattering experiments (including the work of D.M. Pajerowski and the initial stages of work by A. Scheie) used resources at the Spallation Neutron Source, a DOE Office of Science User Facility operated by the Oak Ridge National Laboratory. 
M.G. Gonzalez, B. Bernu, and L. Messio were supported by the French Agence Nationale de la Recherche under Grant No. ANR-18-CE30-0022-04 LINK. 
J. Xing and A. Sefat were supported by U.S. Department of Energy, Basic Energy Sciences, Materials Science and Engineering Division. 
Q.H. and H.D.Z. thank the support from the NSF with Grant No. NSF-DMR-2003117. The NaYbSe$_2$ magnetization measurement performed at NHMFL was supported by Grant No. NSF-DMR-1157490 and the State of Florida.  We are grateful to Tao Xie for discussions on CsYbSe$_2$. 
\end{acknowledgments}

\appendix

\section{Experiments \label{app:Experiments}}

\subsection{Cold Neutron Chopper Spectrometer (CNCS)}

\begin{figure}[h]
	\centering\includegraphics[width=0.15\textwidth]{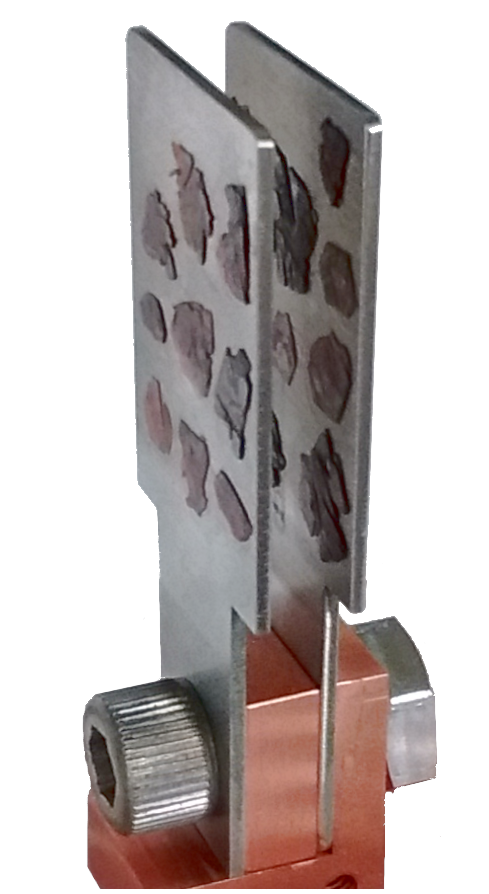}
	\caption{KYbSe$_2$ sample used to measure the field-dependent spin excitations on CNCS. 19 plate-like crystals were coaligned and glued to two aluminum plates with the $[1\bar{1}0]$ direction vertical and the $[001]$ direction orthogonal to the aluminum surface.}
	\label{flo:SampleMount}
\end{figure}

\begin{figure}[h]
	\centering\includegraphics[width=0.28\textwidth]{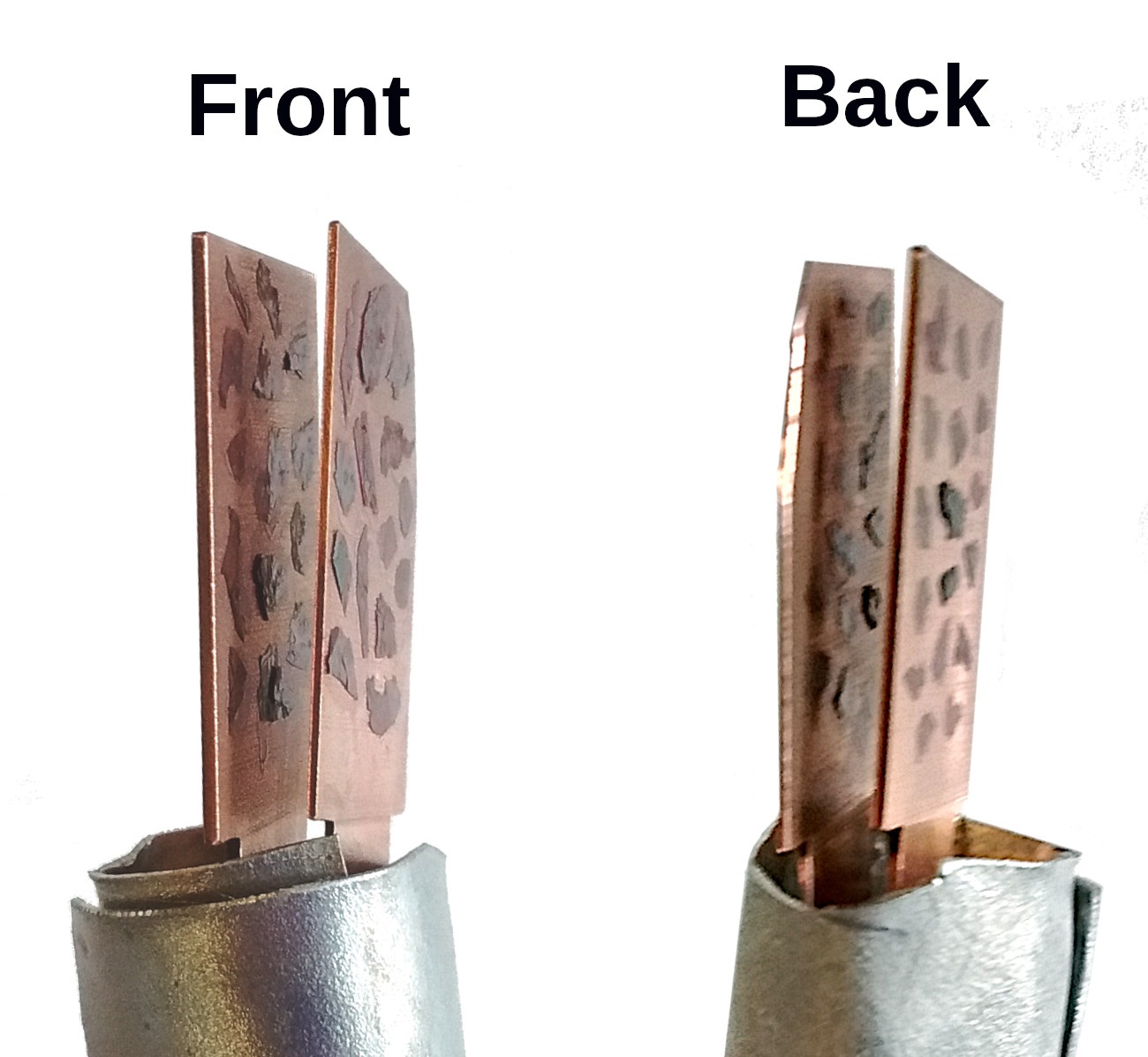}
	\caption{NaYbSe$_2$ sample used to measure the field-dependent spin excitations on CNCS. 61 plate-like crystals were coaligned and glued to two copper plates with the $[1\bar{1}0]$ direction vertical and the $[001]$ direction orthogonal to the copper surface. The grey metal around the bottom is cadmium shielding.}
	\label{flo:NYSSampleMount}
\end{figure}

For the KYbSe$_2$ CNCS experiment, we used a sample made of 19 coaligned crystals (total mass $\sim 200$~mg) oriented in the $(hh\ell)$ scattering plane glued to aluminum plates shown in Fig. \ref{flo:SampleMount}. The sample was mounted in a helium dilution refrigerator with an 8~T magnet and measured with double-disc chopper frequency 300.0 Hz (high-flux mode, 9 degree opening on the double disk) for an incident energy $E_i=3.32$~meV. At each field the sample was rotated 180$^{\circ}$ to map the neutron spectrum. The scattering  was measured at applied fields of 0~T, 4~T, and 8~T at a 100~mK base temperature, with a 12~K zero field data set collected and subtracted as background. 
The same experimental configuration was used for NaYbSe$_2$, but with a sample of 61 coaligned crystals totaling $\sim 300$~mg, glued to copper plates, shown in Fig. \ref{flo:NYSSampleMount}. 

\begin{figure*}
	\centering\includegraphics[width=0.96\textwidth]{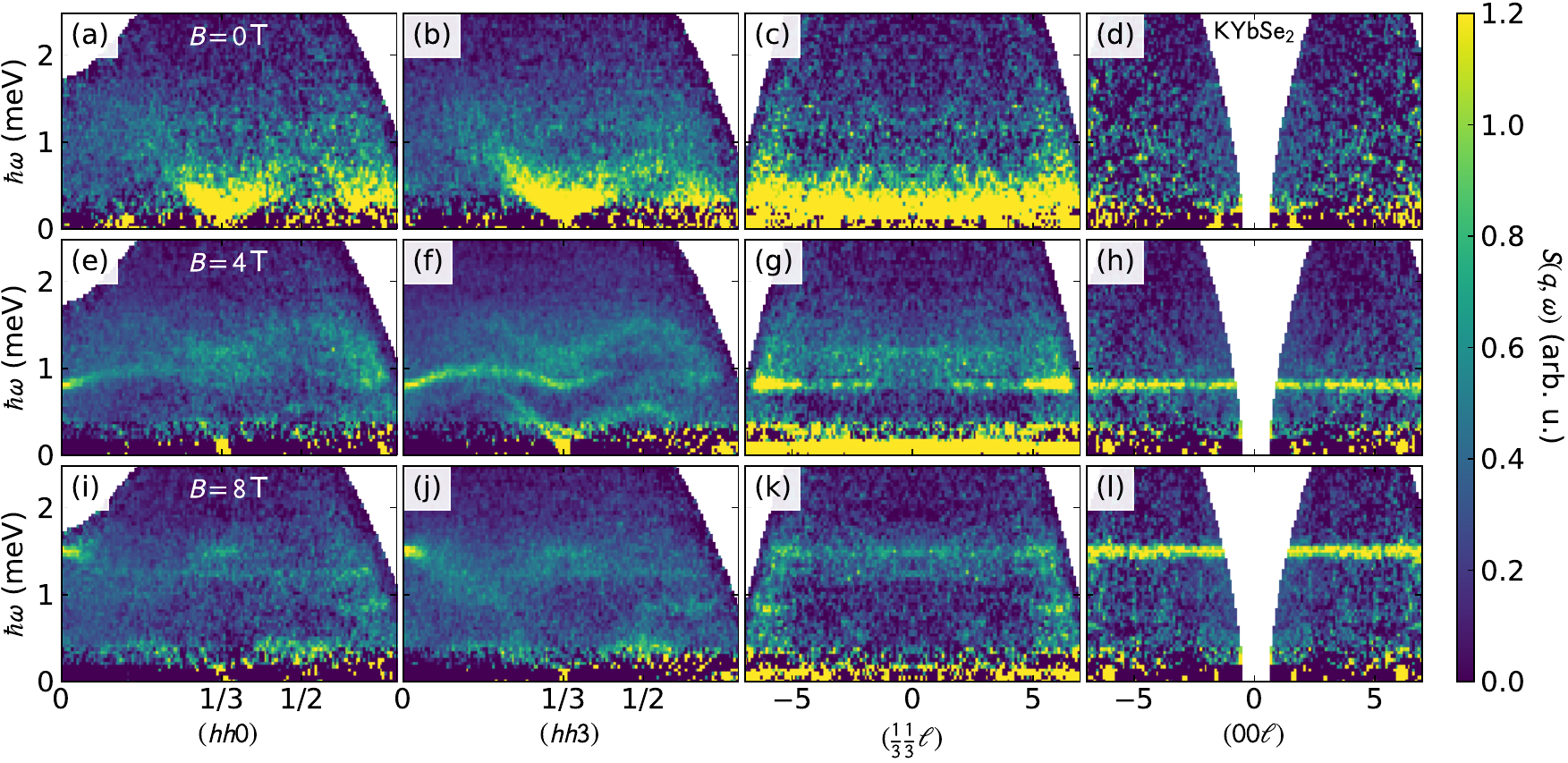}
	\caption{KYbSe$_2$ field dependent scattering without background subtraction. The top row shows the zero field neutron spectrum, the middle row shows the same slices for $B=4$~T, and the bottom row for $B=8$~T. These data should be compared with main text Fig. \ref{flo:Schematic}. At high energy transfers the features are clear, but background subtraction (Fig. \ref{flo:BKG}) is necessary to resolve low energy features.} 
	\label{flo:NeutronScattering_noBG}
\end{figure*}

The processed CNCS data was corrected for the isotropic Yb$^{3+}$ form factor~\cite{BrownFF}. The KYbSe$_2$ data (prior to background subtraction) is shown in Fig. \ref{flo:NeutronScattering_noBG}.  The left two columns are integrated over $\ell \pm 1.5$ reciprocal lattice units (RLU), and the right two columns are integrated over $hh \pm 0.02$~RLU, and the integration ranges for the main text data are the same.

\begin{figure*}
	\centering\includegraphics[width=0.96\textwidth]{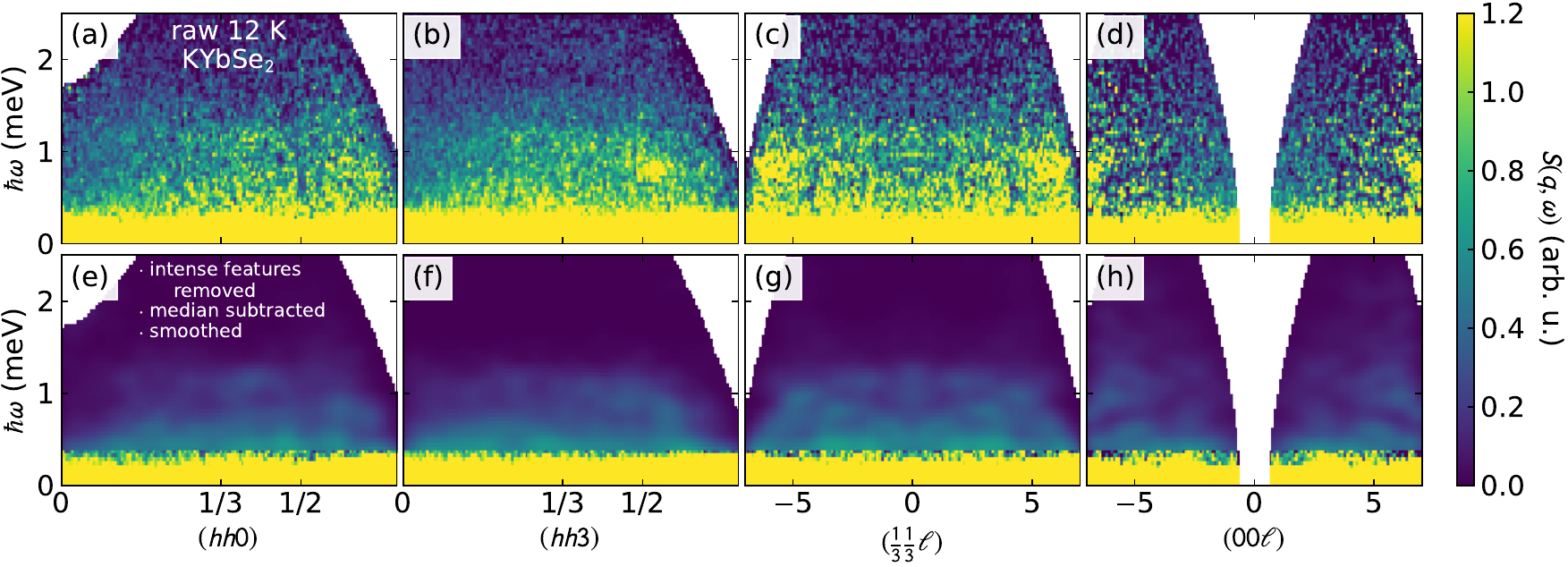}
	\caption{KYbSe$_2$ scattering background. The top row shows the raw  $T=12$~K data. To process this and use it as a background for the data in Fig. \ref{flo:NeutronScattering_noBG}, we removed the intense features [e.g., at 0.7~meV in panel (b)], subtracted the energy-dependent median inelastic scattering intensity from every energy, and convolved with a Gaussian function to smooth the data. This resulted in the background in the bottom row, which was subtracted from the scattering data in Fig. \ref{flo:NeutronScattering_noBG} to generate the data in main text Fig \ref{flo:Schematic}.} 
	\label{flo:BKG}
\end{figure*}

To clearly resolve the low-energy magnetic scattering, it was necessary to subtract the background (sample environment scattering and phonon scattering) from the data.
 Because a perfect background is not available for this data, we created a phenomenological background using 12~K scattering data where magnetic correlations are negligible \cite{scheie2021_KYS,Xing2021_KYS}, shown for KYbSe$_2$ in Fig. \ref{flo:BKG}(a)-(d). Because the higher temperature intensifies phonon scattering, we removed certain intense inelastic features in the slices and filled in via \textit{Astropy's} interpolation routine \cite{robitaille2013astropy}. Then, following the method in Ref. \cite{scheie2021_KYS}, we subtracted the median intensity for each energy from the scattering data for inelastic ($> 0.38$~meV) data. Finally, we convolved the inelastic ($> 0.38$~meV) data with a two-dimensional Gaussian profile to reduce the noise. This final KYbSe$_2$ phenomenological background is shown in Fig. \ref{flo:BKG}(e)-(f). An identical procedure was carried out for the NaYbSe$_2$ background. By comparing Fig. \ref{flo:NeutronScattering_noBG} to the main text Fig. \ref{flo:Schematic}, it is clear that this approach approximately isolates the magnetic scattering quite nicely. 

\subsection{Saturation field for NaYbSe$_2$}

\begin{figure}
	\centering\includegraphics[width=0.45\textwidth]{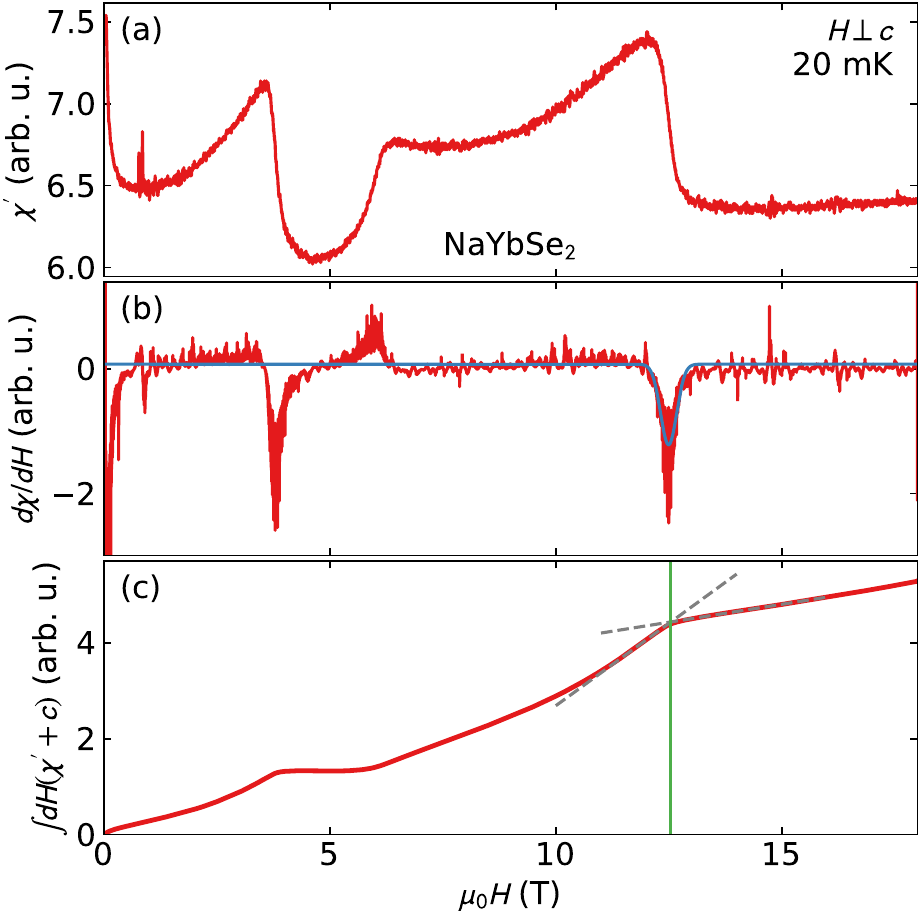}
	\caption{Measured  NaYbSe$_2$ susceptibility at 20~mK. The top panel {\bf a} shows the raw susceptibility, the middle panel {\bf b} shows the first derivative, from which the saturation field is identified as the minimum at 12.49~T, quantified with a fitted Gaussian (blue line). The bottom panel {\bf c} shows the field integral of susceptibility, giving magnetization (up to a linear in $H$ offset). Here the saturation field is identified as the kink, quantified as the intersection of two fitted straight lines to be 12.55~T.}
	\label{flo:NYS_susceptibility}
\end{figure}

We measured the saturation field for NaYbSe$_2$ via low temperature AC susceptibility up to 18~T, as shown in Fig. \ref{flo:NYS_susceptibility}. The saturation transition is visible as a drop in susceptibility. We quantified the saturation field in two ways: first, we used the minimum in the field derivative of susceptibility, fitted with a Gaussian to be at 12.49~T. Second, we used the kink in the field integral of susceptibility, identified as the intersection of two fitted straight lines (one fitted to data above the saturation field, the other fitted to data below it) to be 12.55~T. The mean of these values is 12.52(3)~T, which is taken to be the saturation field and uncertainty for the Hamiltonian fits.

\subsection{Heat capacity uncertainty}

The heat capacity data presented in the main text is made of two different measurements: one measured in a dilution refrigerator \cite{scheie2021_KYS}, and another on a Quantum Design PPMS with a $^3$He insert \cite{Xing2021_KYS}. The KYbSe$_2$ sample mass was measured more precisely for the dilution refrigerator experiment, and thus we scaled the higher temperature PPMS data to match the dilution refrigerator measurement for the temperature window $0.4\>{\rm K} < T < 2.0\>{\rm K}$, giving a renormalizing scale factor of 0.99 to the high temperature data. Heat capacity of nonmagnetic KLuSe$_2$ was subtracted from all data to isolate the magnetic heat capacity.

The uncertainties for the heat capacity values came from three sources: (i) uncertainty in mass normalization, which we estimate from the the difference between the absolute heat capacities of the two different dilution refrigerator measurements on two different KYbSe$_2$ samples reported in Ref. \cite{scheie2021_KYS}. (ii) Variance in heat capacity measured at the same temperature: in the $^3$He heat capacity measurement, data points were repeated three times at each temperature. (iii) Difference in heat capacity between two separate $^3$He refrigerator measurements. The uncertainties from these three sources were added in quadrature to yield the uncertainties plotted in the main text Fig. \ref{flo:HeatCapacity}.

\subsection{AC Calorimetry under pressure}

Heat capacity measurements under hydrostatic pressure in Fig. \ref{flo:HeatCapacityPressure} were performed using a piston-cylinder-type pressure cell with Daphne 7373 oil as the pressure-transmitting medium. The pressure inside the sample space was determined at low temperatures by the shift of the superconducting transition temperature of a piece of Pb. Heat capacity was measured via ac calorimetry technique \cite{Sullivan_1968} using a constantan wire heater and chromel-AuFe0.07\% thermocouple, both attached to the sample using GE varnish. The signal from the thermocouple were amplified using a SR554 low-noise preamplifier transformer and recorded using SR860 lock-in amplifier. Measurements were performed at temperatures down to 85~mK using an adiabatic demagnetization refrigerator (Cambridge Magnetic Refrigeration).

\section{Nonlinear spin waves \label{app:NonLinearSpinWaves}}

\subsection{Fitting the nonlinear spin waves}

The fitted spin wave equation (Eq. \ref{eq:hamiltonian}) contains three unknowns: the magnetic field $h$, the nearest neighbor exchange $J_1$, and the second neighbor exchange $J_2$. Although we know the applied magnetic field in units of Tesla, one has to know the precise $g$-tensor value to convert to the dimensionless (meV) units. Because the fitted $g$-tensors of Yb$^{3+}$ compounds fitted from crystal field excitations have notoriously large uncertainty \cite{scheie2022quantifying}, we avoid using a fitted $g$-tensor to constrain the fit. For KYbSe$_2$, there was enough data points to treat $h$ as a fitted variable, and thus the fit was independent of other measurements. 

For NaYbSe$_2$, with only a single visible mode, we constrained $h$ using the saturation field at low temperatures. In the pure Heisenberg triangular lattice antiferromagnet, the saturation field is $h_{sat} = 3 J_1$. By bulk susceptibility and magnetization, the measured saturation field is $H_{sat} = 12.52(3)$~T. Thus we can convert the applied 4.7~T to dimensionless units: $h = \frac{h_{sat}}{H_{sat}} H = \frac{3 J_1}{12.52(3) \> {\rm T}} 4.7 \> {\rm T} = 1.126(3) \> J_1$. Fortunately, as shown in Fig. \ref{flo:NYS_NLSWfit}, the $J_2/J_1$ ratio is well constrained once the applied field $h$ is defined.

\subsection{Two magnon continuum}

\begin{figure*}
	\centering\includegraphics[width=0.8\textwidth]{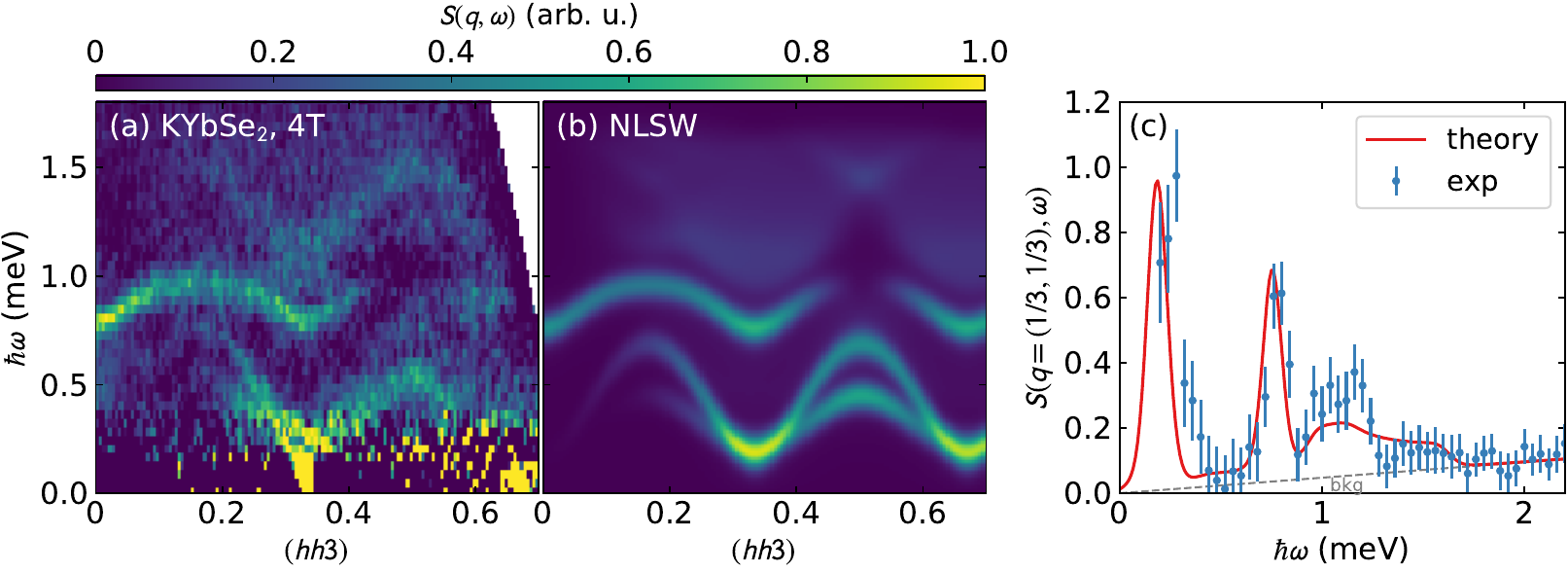}
	\caption{Two magnon continuum of KYbSe$_2$. Panel (a) shows the experimental data at $B=4$~T, and panel (b) shows the nonlinear spin wave calculated intensity with the two-magnon contribution included. Panel (c) compares the experiment with the theory at $(1/3,1/3,3)$. If we assume some small remnant background the overall shape and intensity of the two-magnon contribution is close to what we observe in experiment, though the calculated two-magnon intensity is slightly lower than experiment.}
	\label{flo:twomagnon}
\end{figure*}

In the experimental KYbSe$_2$ scattering, an additional broadened mode is visible above 1~meV which is not captured by the single-magnon NLSWT calculation (main text Fig. \ref{flo:NLSWfit}). However, when we calculate the two-magnon continuum scattering (following the method in Ref. \cite{Kamiya18}), we see a similar mode appear in the simulation, as shown Fig. \ref{flo:twomagnon}(b).

Although the overall character and bandwidth of the calculated two-magnon continuum matches experiment, Fig. \ref{flo:twomagnon}(c) shows the calculated intensity is slightly weaker than what is experimentally observed. 
There could be several explanations for this: (i) the experimental background subtraction was imperfect and left some nonlinear anomalous intensity above 1~meV. (ii) The $g$-tensor is different than the best fit crystal field values ($g_{zz}$ has a very large error bar \cite{scheie2021_KYS}), which leads to different weight being given to $S_{xx}$ vs $S_{yy}$ and $S_{zz}$ and more intensity in the continuum. (iii) The NLSWT approximation neglects higher order effects, and thus it may be slightly underestimating the weight of the continuum scattering (this idea receives support from Bond Operator technique calculations, which show more weight in the continuum \cite{syromyatnikov2022unusual}, and is interpreted in Ref. \cite{xie2022complete} as a two-magnon bound state). Despite this discrepancy, because of the close resemblance to the experimental dispersion, we are confident that the observed mode above 1~meV is in fact a two-magnon scattering effect.

\subsection{Nonlinear effects}

\begin{figure}
	\centering\includegraphics[width=0.4\textwidth]{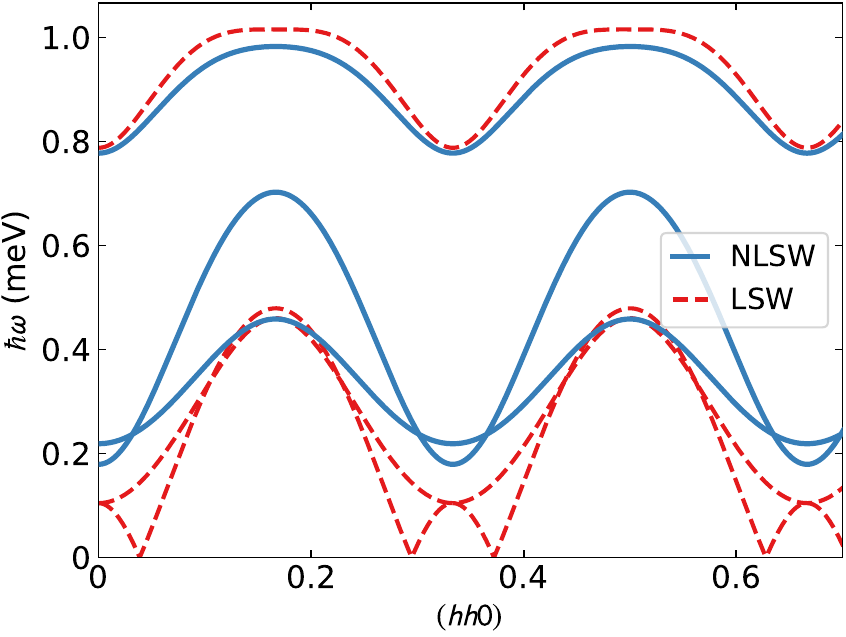}
	\caption{Nonlinear spin wave calculations (blue) compared to linear spin wave calculations (red)  for the triangular lattice using best fit parameters in the main text. Note that at this field, nonlinear corrections to the spin wave modes are significant, especially at low energies.}
	\label{flo:LSW}
\end{figure}

For a highly quantum system like a $S=1/2$ triangular lattice in the 1/3 magnetization plateau phase, nonlinear corrections to magnon dispersions become significant. To illustrate this, Figure \ref{flo:LSW} shows the linear spin wave dispersions compared to the nonlinear dispersions for the best fit KYbSe$_2$ parameters from the main text. Particularly at low energies, the nonlinear corrections are very significant. Note that the existence of a gap (signaling a finite-field plateau phase) requires nonlinear effects to capture; this is because the plateau phase is an inherently quantum-mechanical phenomenon \cite{Kamiya18}. 
Thus accurately extracting the  Hamiltonian parameters from the neutron scattering measurements on KYbSe$_2$ requires a nonlinear spin wave model.

\section{Density functional theory calculations \label{app:DFT}}

We performed the DFT calculations reported in Table \ref{ta:expressure} as implemented in VASP \cite{VASP1,VASP2}. The calculations are performed within the Perdew-Burke-Ernzerhof (PBE) generalized gradient approximation (GGA) \cite{GGA} for the exchange-correlation functional with spin orbit coupling. We use projector augmented wave (PAW) pseudopotentials \cite{PAW1,PAW2} with an energy cutoff of 300 eV and a $9 \times 9 \times 9$ Monkhorst-Pack $k$-point mesh. The pseudopotential for each alkali metal and Yb treated the $s$ and $p$ semi-core states as valence states. 
For calculations with applied hydrostatic pressure 0-5~GPa, the structure was optimized by allowing atomic positions, cell shape, and cell volume to relax until component forces were less than 1 meV/\AA. 
We use Wannier90 \cite{W90,Marzari1997,Souza2001} to create a tight-binding Hamiltonian by projecting the band structure onto real (axial) Yb-$f$ orbitals. The maximal-localization step is not performed in order to maintain the symmetry character of the Wannier functions. The frozen window includes only the seven $f$-bands at the Fermi level which comprise a disconnected manifold of bands; disentanglement is unnecessary. For further details on these calculations, we refer to Ref.~\cite{Villanova2023}. 


\begin{thebibliography}{75}%
	\makeatletter
	\providecommand \@ifxundefined [1]{%
		\@ifx{#1\undefined}
	}%
	\providecommand \@ifnum [1]{%
		\ifnum #1\expandafter \@firstoftwo
		\else \expandafter \@secondoftwo
		\fi
	}%
	\providecommand \@ifx [1]{%
		\ifx #1\expandafter \@firstoftwo
		\else \expandafter \@secondoftwo
		\fi
	}%
	\providecommand \natexlab [1]{#1}%
	\providecommand \enquote  [1]{``#1''}%
	\providecommand \bibnamefont  [1]{#1}%
	\providecommand \bibfnamefont [1]{#1}%
	\providecommand \citenamefont [1]{#1}%
	\providecommand \href@noop [0]{\@secondoftwo}%
	\providecommand \href [0]{\begingroup \@sanitize@url \@href}%
	\providecommand \@href[1]{\@@startlink{#1}\@@href}%
	\providecommand \@@href[1]{\endgroup#1\@@endlink}%
	\providecommand \@sanitize@url [0]{\catcode `\\12\catcode `\$12\catcode
		`\&12\catcode `\#12\catcode `\^12\catcode `\_12\catcode `\%12\relax}%
	\providecommand \@@startlink[1]{}%
	\providecommand \@@endlink[0]{}%
	\providecommand \url  [0]{\begingroup\@sanitize@url \@url }%
	\providecommand \@url [1]{\endgroup\@href {#1}{\urlprefix }}%
	\providecommand \urlprefix  [0]{URL }%
	\providecommand \Eprint [0]{\href }%
	\providecommand \doibase [0]{https://doi.org/}%
	\providecommand \selectlanguage [0]{\@gobble}%
	\providecommand \bibinfo  [0]{\@secondoftwo}%
	\providecommand \bibfield  [0]{\@secondoftwo}%
	\providecommand \translation [1]{[#1]}%
	\providecommand \BibitemOpen [0]{}%
	\providecommand \bibitemStop [0]{}%
	\providecommand \bibitemNoStop [0]{.\EOS\space}%
	\providecommand \EOS [0]{\spacefactor3000\relax}%
	\providecommand \BibitemShut  [1]{\csname bibitem#1\endcsname}%
	\let\auto@bib@innerbib\@empty
	\bibitem [{\citenamefont {Anderson}(1973)}]{Anderson1973}%
	\BibitemOpen
	\bibfield  {author} {\bibinfo {author} {\bibfnamefont {P.}~\bibnamefont
			{Anderson}},\ }\bibfield  {title} {\bibinfo {title} {Resonating valence
			bonds: A new kind of insulator?},\ }\href
	{https://doi.org/https://doi.org/10.1016/0025-5408(73)90167-0} {\bibfield
		{journal} {\bibinfo  {journal} {Materials Research Bulletin}\ }\textbf
		{\bibinfo {volume} {8}},\ \bibinfo {pages} {153 } (\bibinfo {year}
		{1973})}\BibitemShut {NoStop}%
	\bibitem [{\citenamefont {Broholm}\ \emph {et~al.}(2020)\citenamefont
		{Broholm}, \citenamefont {Cava}, \citenamefont {Kivelson}, \citenamefont
		{Nocera}, \citenamefont {Norman},\ and\ \citenamefont
		{Senthil}}]{broholm2019quantum}%
	\BibitemOpen
	\bibfield  {author} {\bibinfo {author} {\bibfnamefont {C.}~\bibnamefont
			{Broholm}}, \bibinfo {author} {\bibfnamefont {R.~J.}\ \bibnamefont {Cava}},
		\bibinfo {author} {\bibfnamefont {S.~A.}\ \bibnamefont {Kivelson}}, \bibinfo
		{author} {\bibfnamefont {D.~G.}\ \bibnamefont {Nocera}}, \bibinfo {author}
		{\bibfnamefont {M.~R.}\ \bibnamefont {Norman}},\ and\ \bibinfo {author}
		{\bibfnamefont {T.}~\bibnamefont {Senthil}},\ }\bibfield  {title} {\bibinfo
		{title} {Quantum spin liquids},\ }\bibfield  {journal} {\bibinfo  {journal}
		{Science}\ }\textbf {\bibinfo {volume} {367}},\ \href
	{https://doi.org/10.1126/science.aay0668} {10.1126/science.aay0668} (\bibinfo
	{year} {2020})\BibitemShut {NoStop}%
	\bibitem [{\citenamefont {Syzranov}\ and\ \citenamefont
		{Ramirez}(2022)}]{Syzranov2022}%
	\BibitemOpen
	\bibfield  {author} {\bibinfo {author} {\bibfnamefont {S.~V.}\ \bibnamefont
			{Syzranov}}\ and\ \bibinfo {author} {\bibfnamefont {A.~P.}\ \bibnamefont
			{Ramirez}},\ }\bibfield  {title} {\bibinfo {title} {Eminuscent phase in
			frustrated magnets: a challenge to quantum spin liquids},\ }\href
	{https://doi.org/10.1038/s41467-022-30739-0} {\bibfield  {journal} {\bibinfo
			{journal} {Nature Communications}\ }\textbf {\bibinfo {volume} {13}},\
		\bibinfo {pages} {2993} (\bibinfo {year} {2022})}\BibitemShut {NoStop}%
	\bibitem [{\citenamefont {Balents}(2010)}]{Balents2010review}%
	\BibitemOpen
	\bibfield  {author} {\bibinfo {author} {\bibfnamefont {L.}~\bibnamefont
			{Balents}},\ }\bibfield  {title} {\bibinfo {title} {Spin liquids in
			frustrated magnets},\ }\href {https://doi.org/10.1038/nature08917} {\bibfield
		{journal} {\bibinfo  {journal} {Nature}\ }\textbf {\bibinfo {volume}
			{464}},\ \bibinfo {pages} {199} (\bibinfo {year} {2010})}\BibitemShut
	{NoStop}%
	\bibitem [{\citenamefont {Zhou}\ \emph {et~al.}(2017)\citenamefont {Zhou},
		\citenamefont {Kanoda},\ and\ \citenamefont {Ng}}]{Zhou_2017_Quantum}%
	\BibitemOpen
	\bibfield  {author} {\bibinfo {author} {\bibfnamefont {Y.}~\bibnamefont
			{Zhou}}, \bibinfo {author} {\bibfnamefont {K.}~\bibnamefont {Kanoda}},\ and\
		\bibinfo {author} {\bibfnamefont {T.-K.}\ \bibnamefont {Ng}},\ }\bibfield
	{title} {\bibinfo {title} {Quantum spin liquid states},\ }\href
	{https://doi.org/10.1103/RevModPhys.89.025003} {\bibfield  {journal}
		{\bibinfo  {journal} {Rev. Mod. Phys.}\ }\textbf {\bibinfo {volume} {89}},\
		\bibinfo {pages} {025003} (\bibinfo {year} {2017})}\BibitemShut {NoStop}%
	\bibitem [{\citenamefont {Nayak}\ \emph {et~al.}(2008)\citenamefont {Nayak},
		\citenamefont {Simon}, \citenamefont {Stern}, \citenamefont {Freedman},\ and\
		\citenamefont {Das~Sarma}}]{Nayak_2008}%
	\BibitemOpen
	\bibfield  {author} {\bibinfo {author} {\bibfnamefont {C.}~\bibnamefont
			{Nayak}}, \bibinfo {author} {\bibfnamefont {S.~H.}\ \bibnamefont {Simon}},
		\bibinfo {author} {\bibfnamefont {A.}~\bibnamefont {Stern}}, \bibinfo
		{author} {\bibfnamefont {M.}~\bibnamefont {Freedman}},\ and\ \bibinfo
		{author} {\bibfnamefont {S.}~\bibnamefont {Das~Sarma}},\ }\bibfield  {title}
	{\bibinfo {title} {Non-abelian anyons and topological quantum computation},\
	}\href {https://doi.org/10.1103/RevModPhys.80.1083} {\bibfield  {journal}
		{\bibinfo  {journal} {Rev. Mod. Phys.}\ }\textbf {\bibinfo {volume} {80}},\
		\bibinfo {pages} {1083} (\bibinfo {year} {2008})}\BibitemShut {NoStop}%
	\bibitem [{\citenamefont {Tokura}\ \emph {et~al.}(2017)\citenamefont {Tokura},
		\citenamefont {Kawasaki},\ and\ \citenamefont
		{Nagaosa}}]{tokura2017emergent}%
	\BibitemOpen
	\bibfield  {author} {\bibinfo {author} {\bibfnamefont {Y.}~\bibnamefont
			{Tokura}}, \bibinfo {author} {\bibfnamefont {M.}~\bibnamefont {Kawasaki}},\
		and\ \bibinfo {author} {\bibfnamefont {N.}~\bibnamefont {Nagaosa}},\
	}\bibfield  {title} {\bibinfo {title} {Emergent functions of quantum
			materials},\ }\href {https://doi.org/10.1038/nphys4274} {\bibfield  {journal}
		{\bibinfo  {journal} {Nature Physics}\ }\textbf {\bibinfo {volume} {13}},\
		\bibinfo {pages} {1056} (\bibinfo {year} {2017})}\BibitemShut {NoStop}%
	\bibitem [{\citenamefont {Capriotti}\ \emph {et~al.}(1999)\citenamefont
		{Capriotti}, \citenamefont {Trumper},\ and\ \citenamefont
		{Sorella}}]{Capriotti_1999}%
	\BibitemOpen
	\bibfield  {author} {\bibinfo {author} {\bibfnamefont {L.}~\bibnamefont
			{Capriotti}}, \bibinfo {author} {\bibfnamefont {A.~E.}\ \bibnamefont
			{Trumper}},\ and\ \bibinfo {author} {\bibfnamefont {S.}~\bibnamefont
			{Sorella}},\ }\bibfield  {title} {\bibinfo {title} {Long-range n\'eel order
			in the triangular heisenberg model},\ }\href
	{https://doi.org/10.1103/PhysRevLett.82.3899} {\bibfield  {journal} {\bibinfo
			{journal} {Phys. Rev. Lett.}\ }\textbf {\bibinfo {volume} {82}},\ \bibinfo
		{pages} {3899} (\bibinfo {year} {1999})}\BibitemShut {NoStop}%
	\bibitem [{\citenamefont {White}\ and\ \citenamefont
		{Chernyshev}(2007)}]{White_2007}%
	\BibitemOpen
	\bibfield  {author} {\bibinfo {author} {\bibfnamefont {S.~R.}\ \bibnamefont
			{White}}\ and\ \bibinfo {author} {\bibfnamefont {A.~L.}\ \bibnamefont
			{Chernyshev}},\ }\bibfield  {title} {\bibinfo {title} {Ne\'el order in square
			and triangular lattice heisenberg models},\ }\href
	{https://doi.org/10.1103/PhysRevLett.99.127004} {\bibfield  {journal}
		{\bibinfo  {journal} {Phys. Rev. Lett.}\ }\textbf {\bibinfo {volume} {99}},\
		\bibinfo {pages} {127004} (\bibinfo {year} {2007})}\BibitemShut {NoStop}%
	\bibitem [{\citenamefont {Zhu}\ and\ \citenamefont
		{White}(2015)}]{PhysRevB.92.041105}%
	\BibitemOpen
	\bibfield  {author} {\bibinfo {author} {\bibfnamefont {Z.}~\bibnamefont
			{Zhu}}\ and\ \bibinfo {author} {\bibfnamefont {S.~R.}\ \bibnamefont
			{White}},\ }\bibfield  {title} {\bibinfo {title} {Spin liquid phase of the
			$s=\frac{1}{2}\phantom{\rule{4.pt}{0ex}}{J}_{1}\ensuremath{-}{J}_{2}$
			heisenberg model on the triangular lattice},\ }\href
	{https://doi.org/10.1103/PhysRevB.92.041105} {\bibfield  {journal} {\bibinfo
			{journal} {Phys. Rev. B}\ }\textbf {\bibinfo {volume} {92}},\ \bibinfo
		{pages} {041105} (\bibinfo {year} {2015})}\BibitemShut {NoStop}%
	\bibitem [{\citenamefont {Hu}\ \emph {et~al.}(2015)\citenamefont {Hu},
		\citenamefont {Gong}, \citenamefont {Zhu},\ and\ \citenamefont
		{Sheng}}]{PhysRevB.92.140403}%
	\BibitemOpen
	\bibfield  {author} {\bibinfo {author} {\bibfnamefont {W.-J.}\ \bibnamefont
			{Hu}}, \bibinfo {author} {\bibfnamefont {S.-S.}\ \bibnamefont {Gong}},
		\bibinfo {author} {\bibfnamefont {W.}~\bibnamefont {Zhu}},\ and\ \bibinfo
		{author} {\bibfnamefont {D.~N.}\ \bibnamefont {Sheng}},\ }\bibfield  {title}
	{\bibinfo {title} {Competing spin-liquid states in the spin-$\frac{1}{2}$
			heisenberg model on the triangular lattice},\ }\href
	{https://doi.org/10.1103/PhysRevB.92.140403} {\bibfield  {journal} {\bibinfo
			{journal} {Phys. Rev. B}\ }\textbf {\bibinfo {volume} {92}},\ \bibinfo
		{pages} {140403} (\bibinfo {year} {2015})}\BibitemShut {NoStop}%
	\bibitem [{\citenamefont {Iqbal}\ \emph {et~al.}(2016)\citenamefont {Iqbal},
		\citenamefont {Hu}, \citenamefont {Thomale}, \citenamefont {Poilblanc},\ and\
		\citenamefont {Becca}}]{PhysRevB.93.144411}%
	\BibitemOpen
	\bibfield  {author} {\bibinfo {author} {\bibfnamefont {Y.}~\bibnamefont
			{Iqbal}}, \bibinfo {author} {\bibfnamefont {W.-J.}\ \bibnamefont {Hu}},
		\bibinfo {author} {\bibfnamefont {R.}~\bibnamefont {Thomale}}, \bibinfo
		{author} {\bibfnamefont {D.}~\bibnamefont {Poilblanc}},\ and\ \bibinfo
		{author} {\bibfnamefont {F.}~\bibnamefont {Becca}},\ }\bibfield  {title}
	{\bibinfo {title} {Spin liquid nature in the heisenberg
			${J}_{1}\ensuremath{-}{J}_{2}$ triangular antiferromagnet},\ }\href
	{https://doi.org/10.1103/PhysRevB.93.144411} {\bibfield  {journal} {\bibinfo
			{journal} {Phys. Rev. B}\ }\textbf {\bibinfo {volume} {93}},\ \bibinfo
		{pages} {144411} (\bibinfo {year} {2016})}\BibitemShut {NoStop}%
	\bibitem [{\citenamefont {Saadatmand}\ and\ \citenamefont
		{McCulloch}(2016)}]{PhysRevB.94.121111}%
	\BibitemOpen
	\bibfield  {author} {\bibinfo {author} {\bibfnamefont {S.~N.}\ \bibnamefont
			{Saadatmand}}\ and\ \bibinfo {author} {\bibfnamefont {I.~P.}\ \bibnamefont
			{McCulloch}},\ }\bibfield  {title} {\bibinfo {title} {Symmetry
			fractionalization in the topological phase of the spin-$\frac{1}{2}$
			${J}_{1}\text{\ensuremath{-}}{J}_{2}$ triangular heisenberg model},\ }\href
	{https://doi.org/10.1103/PhysRevB.94.121111} {\bibfield  {journal} {\bibinfo
			{journal} {Phys. Rev. B}\ }\textbf {\bibinfo {volume} {94}},\ \bibinfo
		{pages} {121111} (\bibinfo {year} {2016})}\BibitemShut {NoStop}%
	\bibitem [{\citenamefont {Wietek}\ and\ \citenamefont
		{L\"auchli}(2017)}]{PhysRevB.95.035141}%
	\BibitemOpen
	\bibfield  {author} {\bibinfo {author} {\bibfnamefont {A.}~\bibnamefont
			{Wietek}}\ and\ \bibinfo {author} {\bibfnamefont {A.~M.}\ \bibnamefont
			{L\"auchli}},\ }\bibfield  {title} {\bibinfo {title} {Chiral spin liquid and
			quantum criticality in extended $s=\frac{1}{2}$ heisenberg models on the
			triangular lattice},\ }\href {https://doi.org/10.1103/PhysRevB.95.035141}
	{\bibfield  {journal} {\bibinfo  {journal} {Phys. Rev. B}\ }\textbf {\bibinfo
			{volume} {95}},\ \bibinfo {pages} {035141} (\bibinfo {year}
		{2017})}\BibitemShut {NoStop}%
	\bibitem [{\citenamefont {Gong}\ \emph {et~al.}(2017)\citenamefont {Gong},
		\citenamefont {Zhu}, \citenamefont {Zhu}, \citenamefont {Sheng},\ and\
		\citenamefont {Yang}}]{PhysRevB.96.075116}%
	\BibitemOpen
	\bibfield  {author} {\bibinfo {author} {\bibfnamefont {S.-S.}\ \bibnamefont
			{Gong}}, \bibinfo {author} {\bibfnamefont {W.}~\bibnamefont {Zhu}}, \bibinfo
		{author} {\bibfnamefont {J.-X.}\ \bibnamefont {Zhu}}, \bibinfo {author}
		{\bibfnamefont {D.~N.}\ \bibnamefont {Sheng}},\ and\ \bibinfo {author}
		{\bibfnamefont {K.}~\bibnamefont {Yang}},\ }\bibfield  {title} {\bibinfo
		{title} {Global phase diagram and quantum spin liquids in a
			spin-$\frac{1}{2}$ triangular antiferromagnet},\ }\href
	{https://doi.org/10.1103/PhysRevB.96.075116} {\bibfield  {journal} {\bibinfo
			{journal} {Phys. Rev. B}\ }\textbf {\bibinfo {volume} {96}},\ \bibinfo
		{pages} {075116} (\bibinfo {year} {2017})}\BibitemShut {NoStop}%
	\bibitem [{\citenamefont {Hu}\ \emph {et~al.}(2019)\citenamefont {Hu},
		\citenamefont {Zhu}, \citenamefont {Eggert},\ and\ \citenamefont
		{He}}]{PhysRevLett.123.207203}%
	\BibitemOpen
	\bibfield  {author} {\bibinfo {author} {\bibfnamefont {S.}~\bibnamefont
			{Hu}}, \bibinfo {author} {\bibfnamefont {W.}~\bibnamefont {Zhu}}, \bibinfo
		{author} {\bibfnamefont {S.}~\bibnamefont {Eggert}},\ and\ \bibinfo {author}
		{\bibfnamefont {Y.-C.}\ \bibnamefont {He}},\ }\bibfield  {title} {\bibinfo
		{title} {Dirac spin liquid on the spin-$1/2$ triangular heisenberg
			antiferromagnet},\ }\href {https://doi.org/10.1103/PhysRevLett.123.207203}
	{\bibfield  {journal} {\bibinfo  {journal} {Phys. Rev. Lett.}\ }\textbf
		{\bibinfo {volume} {123}},\ \bibinfo {pages} {207203} (\bibinfo {year}
		{2019})}\BibitemShut {NoStop}%
	\bibitem [{\citenamefont {Zhu}\ \emph {et~al.}(2018)\citenamefont {Zhu},
		\citenamefont {Maksimov}, \citenamefont {White},\ and\ \citenamefont
		{Chernyshev}}]{Zhu_2018}%
	\BibitemOpen
	\bibfield  {author} {\bibinfo {author} {\bibfnamefont {Z.}~\bibnamefont
			{Zhu}}, \bibinfo {author} {\bibfnamefont {P.~A.}\ \bibnamefont {Maksimov}},
		\bibinfo {author} {\bibfnamefont {S.~R.}\ \bibnamefont {White}},\ and\
		\bibinfo {author} {\bibfnamefont {A.~L.}\ \bibnamefont {Chernyshev}},\
	}\bibfield  {title} {\bibinfo {title} {Topography of spin liquids on a
			triangular lattice},\ }\href {https://doi.org/10.1103/PhysRevLett.120.207203}
	{\bibfield  {journal} {\bibinfo  {journal} {Phys. Rev. Lett.}\ }\textbf
		{\bibinfo {volume} {120}},\ \bibinfo {pages} {207203} (\bibinfo {year}
		{2018})}\BibitemShut {NoStop}%
	\bibitem [{\citenamefont {Gong}\ \emph {et~al.}(2019)\citenamefont {Gong},
		\citenamefont {Zheng}, \citenamefont {Lee}, \citenamefont {Lu},\ and\
		\citenamefont {Sheng}}]{PhysRevB.100.241111}%
	\BibitemOpen
	\bibfield  {author} {\bibinfo {author} {\bibfnamefont {S.-S.}\ \bibnamefont
			{Gong}}, \bibinfo {author} {\bibfnamefont {W.}~\bibnamefont {Zheng}},
		\bibinfo {author} {\bibfnamefont {M.}~\bibnamefont {Lee}}, \bibinfo {author}
		{\bibfnamefont {Y.-M.}\ \bibnamefont {Lu}},\ and\ \bibinfo {author}
		{\bibfnamefont {D.~N.}\ \bibnamefont {Sheng}},\ }\bibfield  {title} {\bibinfo
		{title} {Chiral spin liquid with spinon fermi surfaces in the
			spin-$\frac{1}{2}$ triangular heisenberg model},\ }\href
	{https://doi.org/10.1103/PhysRevB.100.241111} {\bibfield  {journal} {\bibinfo
			{journal} {Phys. Rev. B}\ }\textbf {\bibinfo {volume} {100}},\ \bibinfo
		{pages} {241111} (\bibinfo {year} {2019})}\BibitemShut {NoStop}%
	\bibitem [{\citenamefont {Ito}\ \emph {et~al.}(2017)\citenamefont {Ito},
		\citenamefont {Kurita}, \citenamefont {Tanaka}, \citenamefont
		{Ohira-Kawamura}, \citenamefont {Nakajima}, \citenamefont {Itoh},
		\citenamefont {Kuwahara},\ and\ \citenamefont {Kakurai}}]{Ito2017}%
	\BibitemOpen
	\bibfield  {author} {\bibinfo {author} {\bibfnamefont {S.}~\bibnamefont
			{Ito}}, \bibinfo {author} {\bibfnamefont {N.}~\bibnamefont {Kurita}},
		\bibinfo {author} {\bibfnamefont {H.}~\bibnamefont {Tanaka}}, \bibinfo
		{author} {\bibfnamefont {S.}~\bibnamefont {Ohira-Kawamura}}, \bibinfo
		{author} {\bibfnamefont {K.}~\bibnamefont {Nakajima}}, \bibinfo {author}
		{\bibfnamefont {S.}~\bibnamefont {Itoh}}, \bibinfo {author} {\bibfnamefont
			{K.}~\bibnamefont {Kuwahara}},\ and\ \bibinfo {author} {\bibfnamefont
			{K.}~\bibnamefont {Kakurai}},\ }\bibfield  {title} {\bibinfo {title}
		{Structure of the magnetic excitations in the spin-1/2 triangular-lattice
			heisenberg antiferromagnet
			{${\mathrm{Ba}}_{3}{\mathrm{CoSb}}_{2}{\mathbf{O}}_{9}$}},\ }\href
	{https://doi.org/10.1038/s41467-017-00316-x} {\bibfield  {journal} {\bibinfo
			{journal} {Nature Communications}\ }\textbf {\bibinfo {volume} {8}},\
		\bibinfo {pages} {235} (\bibinfo {year} {2017})}\BibitemShut {NoStop}%
	\bibitem [{\citenamefont {Macdougal}\ \emph {et~al.}(2020)\citenamefont
		{Macdougal}, \citenamefont {Williams}, \citenamefont {Prabhakaran},
		\citenamefont {Bewley}, \citenamefont {Voneshen},\ and\ \citenamefont
		{Coldea}}]{Macdougal_2020}%
	\BibitemOpen
	\bibfield  {author} {\bibinfo {author} {\bibfnamefont {D.}~\bibnamefont
			{Macdougal}}, \bibinfo {author} {\bibfnamefont {S.}~\bibnamefont {Williams}},
		\bibinfo {author} {\bibfnamefont {D.}~\bibnamefont {Prabhakaran}}, \bibinfo
		{author} {\bibfnamefont {R.~I.}\ \bibnamefont {Bewley}}, \bibinfo {author}
		{\bibfnamefont {D.~J.}\ \bibnamefont {Voneshen}},\ and\ \bibinfo {author}
		{\bibfnamefont {R.}~\bibnamefont {Coldea}},\ }\bibfield  {title} {\bibinfo
		{title} {Avoided quasiparticle decay and enhanced excitation continuum in the
			spin-$\frac{1}{2}$ near-heisenberg triangular antiferromagnet
			{${\mathrm{Ba}}_{3}{\mathrm{CoSb}}_{2}{\mathrm{O}}_{9}$}},\ }\href
	{https://doi.org/10.1103/PhysRevB.102.064421} {\bibfield  {journal} {\bibinfo
			{journal} {Phys. Rev. B}\ }\textbf {\bibinfo {volume} {102}},\ \bibinfo
		{pages} {064421} (\bibinfo {year} {2020})}\BibitemShut {NoStop}%
	\bibitem [{\citenamefont {Ma}\ \emph {et~al.}(2016)\citenamefont {Ma},
		\citenamefont {Kamiya}, \citenamefont {Hong}, \citenamefont {Cao},
		\citenamefont {Ehlers}, \citenamefont {Tian}, \citenamefont {Batista},
		\citenamefont {Dun}, \citenamefont {Zhou},\ and\ \citenamefont
		{Matsuda}}]{Ma_2016}%
	\BibitemOpen
	\bibfield  {author} {\bibinfo {author} {\bibfnamefont {J.}~\bibnamefont
			{Ma}}, \bibinfo {author} {\bibfnamefont {Y.}~\bibnamefont {Kamiya}}, \bibinfo
		{author} {\bibfnamefont {T.}~\bibnamefont {Hong}}, \bibinfo {author}
		{\bibfnamefont {H.~B.}\ \bibnamefont {Cao}}, \bibinfo {author} {\bibfnamefont
			{G.}~\bibnamefont {Ehlers}}, \bibinfo {author} {\bibfnamefont
			{W.}~\bibnamefont {Tian}}, \bibinfo {author} {\bibfnamefont {C.~D.}\
			\bibnamefont {Batista}}, \bibinfo {author} {\bibfnamefont {Z.~L.}\
			\bibnamefont {Dun}}, \bibinfo {author} {\bibfnamefont {H.~D.}\ \bibnamefont
			{Zhou}},\ and\ \bibinfo {author} {\bibfnamefont {M.}~\bibnamefont
			{Matsuda}},\ }\bibfield  {title} {\bibinfo {title} {Static and dynamical
			properties of the spin-$1/2$ equilateral triangular-lattice antiferromagnet
			{${\mathrm{Ba}}_{3}{\mathrm{CoSb}}_{2}{\mathrm{O}}_{9}$}},\ }\href
	{https://doi.org/10.1103/PhysRevLett.116.087201} {\bibfield  {journal}
		{\bibinfo  {journal} {Phys. Rev. Lett.}\ }\textbf {\bibinfo {volume} {116}},\
		\bibinfo {pages} {087201} (\bibinfo {year} {2016})}\BibitemShut {NoStop}%
	\bibitem [{\citenamefont {Li}\ \emph {et~al.}(2015)\citenamefont {Li},
		\citenamefont {Chen}, \citenamefont {Tong}, \citenamefont {Pi}, \citenamefont
		{Liu}, \citenamefont {Yang}, \citenamefont {Wang},\ and\ \citenamefont
		{Zhang}}]{Li_2015_YMGO}%
	\BibitemOpen
	\bibfield  {author} {\bibinfo {author} {\bibfnamefont {Y.}~\bibnamefont
			{Li}}, \bibinfo {author} {\bibfnamefont {G.}~\bibnamefont {Chen}}, \bibinfo
		{author} {\bibfnamefont {W.}~\bibnamefont {Tong}}, \bibinfo {author}
		{\bibfnamefont {L.}~\bibnamefont {Pi}}, \bibinfo {author} {\bibfnamefont
			{J.}~\bibnamefont {Liu}}, \bibinfo {author} {\bibfnamefont {Z.}~\bibnamefont
			{Yang}}, \bibinfo {author} {\bibfnamefont {X.}~\bibnamefont {Wang}},\ and\
		\bibinfo {author} {\bibfnamefont {Q.}~\bibnamefont {Zhang}},\ }\bibfield
	{title} {\bibinfo {title} {Rare-earth triangular lattice spin liquid: A
			single-crystal study of {${\mathrm{YbMgGaO}}_{4}$}},\ }\href
	{https://doi.org/10.1103/PhysRevLett.115.167203} {\bibfield  {journal}
		{\bibinfo  {journal} {Phys. Rev. Lett.}\ }\textbf {\bibinfo {volume} {115}},\
		\bibinfo {pages} {167203} (\bibinfo {year} {2015})}\BibitemShut {NoStop}%
	\bibitem [{\citenamefont {Shen}\ \emph {et~al.}(2016)\citenamefont {Shen},
		\citenamefont {Li}, \citenamefont {Wo}, \citenamefont {Li}, \citenamefont
		{Shen}, \citenamefont {Pan}, \citenamefont {Wang}, \citenamefont {Walker},
		\citenamefont {Steffens}, \citenamefont {Boehm}, \citenamefont {Hao},
		\citenamefont {Quintero-Castro}, \citenamefont {Harriger}, \citenamefont
		{Frontzek}, \citenamefont {Hao}, \citenamefont {Meng}, \citenamefont {Zhang},
		\citenamefont {Chen},\ and\ \citenamefont {Zhao}}]{Shen2016}%
	\BibitemOpen
	\bibfield  {author} {\bibinfo {author} {\bibfnamefont {Y.}~\bibnamefont
			{Shen}}, \bibinfo {author} {\bibfnamefont {Y.-D.}\ \bibnamefont {Li}},
		\bibinfo {author} {\bibfnamefont {H.}~\bibnamefont {Wo}}, \bibinfo {author}
		{\bibfnamefont {Y.}~\bibnamefont {Li}}, \bibinfo {author} {\bibfnamefont
			{S.}~\bibnamefont {Shen}}, \bibinfo {author} {\bibfnamefont {B.}~\bibnamefont
			{Pan}}, \bibinfo {author} {\bibfnamefont {Q.}~\bibnamefont {Wang}}, \bibinfo
		{author} {\bibfnamefont {H.~C.}\ \bibnamefont {Walker}}, \bibinfo {author}
		{\bibfnamefont {P.}~\bibnamefont {Steffens}}, \bibinfo {author}
		{\bibfnamefont {M.}~\bibnamefont {Boehm}}, \bibinfo {author} {\bibfnamefont
			{Y.}~\bibnamefont {Hao}}, \bibinfo {author} {\bibfnamefont {D.~L.}\
			\bibnamefont {Quintero-Castro}}, \bibinfo {author} {\bibfnamefont {L.~W.}\
			\bibnamefont {Harriger}}, \bibinfo {author} {\bibfnamefont {M.~D.}\
			\bibnamefont {Frontzek}}, \bibinfo {author} {\bibfnamefont {L.}~\bibnamefont
			{Hao}}, \bibinfo {author} {\bibfnamefont {S.}~\bibnamefont {Meng}}, \bibinfo
		{author} {\bibfnamefont {Q.}~\bibnamefont {Zhang}}, \bibinfo {author}
		{\bibfnamefont {G.}~\bibnamefont {Chen}},\ and\ \bibinfo {author}
		{\bibfnamefont {J.}~\bibnamefont {Zhao}},\ }\bibfield  {title} {\bibinfo
		{title} {Evidence for a spinon fermi surface in a triangular-lattice
			quantum-spin-liquid candidate},\ }\href {https://doi.org/10.1038/nature20614}
	{\bibfield  {journal} {\bibinfo  {journal} {Nature}\ }\textbf {\bibinfo
			{volume} {540}},\ \bibinfo {pages} {559} (\bibinfo {year}
		{2016})}\BibitemShut {NoStop}%
	\bibitem [{\citenamefont {Paddison}\ \emph {et~al.}(2017)\citenamefont
		{Paddison}, \citenamefont {Daum}, \citenamefont {Dun}, \citenamefont
		{Ehlers}, \citenamefont {Liu}, \citenamefont {Stone}, \citenamefont {Zhou},\
		and\ \citenamefont {Mourigal}}]{Paddison2017}%
	\BibitemOpen
	\bibfield  {author} {\bibinfo {author} {\bibfnamefont {J.~A.~M.}\
			\bibnamefont {Paddison}}, \bibinfo {author} {\bibfnamefont {M.}~\bibnamefont
			{Daum}}, \bibinfo {author} {\bibfnamefont {Z.}~\bibnamefont {Dun}}, \bibinfo
		{author} {\bibfnamefont {G.}~\bibnamefont {Ehlers}}, \bibinfo {author}
		{\bibfnamefont {Y.}~\bibnamefont {Liu}}, \bibinfo {author} {\bibfnamefont
			{M.}~\bibnamefont {Stone}}, \bibinfo {author} {\bibfnamefont
			{H.}~\bibnamefont {Zhou}},\ and\ \bibinfo {author} {\bibfnamefont
			{M.}~\bibnamefont {Mourigal}},\ }\bibfield  {title} {\bibinfo {title}
		{Continuous excitations of the triangular-lattice quantum spin liquid
			{${\mathrm{YbMgGaO}}_{4}$}},\ }\href {https://doi.org/10.1038/nphys3971}
	{\bibfield  {journal} {\bibinfo  {journal} {Nature Physics}\ }\textbf
		{\bibinfo {volume} {13}},\ \bibinfo {pages} {117} (\bibinfo {year}
		{2017})}\BibitemShut {NoStop}%
	\bibitem [{\citenamefont {Xu}\ \emph {et~al.}(2016)\citenamefont {Xu},
		\citenamefont {Zhang}, \citenamefont {Li}, \citenamefont {Yu}, \citenamefont
		{Hong}, \citenamefont {Zhang},\ and\ \citenamefont {Li}}]{Xu_2016_YMGO}%
	\BibitemOpen
	\bibfield  {author} {\bibinfo {author} {\bibfnamefont {Y.}~\bibnamefont
			{Xu}}, \bibinfo {author} {\bibfnamefont {J.}~\bibnamefont {Zhang}}, \bibinfo
		{author} {\bibfnamefont {Y.~S.}\ \bibnamefont {Li}}, \bibinfo {author}
		{\bibfnamefont {Y.~J.}\ \bibnamefont {Yu}}, \bibinfo {author} {\bibfnamefont
			{X.~C.}\ \bibnamefont {Hong}}, \bibinfo {author} {\bibfnamefont {Q.~M.}\
			\bibnamefont {Zhang}},\ and\ \bibinfo {author} {\bibfnamefont {S.~Y.}\
			\bibnamefont {Li}},\ }\bibfield  {title} {\bibinfo {title} {Absence of
			magnetic thermal conductivity in the quantum spin-liquid candidate
			{${\mathrm{YbMgGaO}}_{4}$}},\ }\href
	{https://doi.org/10.1103/PhysRevLett.117.267202} {\bibfield  {journal}
		{\bibinfo  {journal} {Phys. Rev. Lett.}\ }\textbf {\bibinfo {volume} {117}},\
		\bibinfo {pages} {267202} (\bibinfo {year} {2016})}\BibitemShut {NoStop}%
	\bibitem [{\citenamefont {Zhu}\ \emph {et~al.}(2017)\citenamefont {Zhu},
		\citenamefont {Maksimov}, \citenamefont {White},\ and\ \citenamefont
		{Chernyshev}}]{Zhu_2017_YMGO}%
	\BibitemOpen
	\bibfield  {author} {\bibinfo {author} {\bibfnamefont {Z.}~\bibnamefont
			{Zhu}}, \bibinfo {author} {\bibfnamefont {P.~A.}\ \bibnamefont {Maksimov}},
		\bibinfo {author} {\bibfnamefont {S.~R.}\ \bibnamefont {White}},\ and\
		\bibinfo {author} {\bibfnamefont {A.~L.}\ \bibnamefont {Chernyshev}},\
	}\bibfield  {title} {\bibinfo {title} {Disorder-induced mimicry of a spin
			liquid in {${\mathrm{YbMgGaO}}_{4}$}},\ }\href
	{https://doi.org/10.1103/PhysRevLett.119.157201} {\bibfield  {journal}
		{\bibinfo  {journal} {Phys. Rev. Lett.}\ }\textbf {\bibinfo {volume} {119}},\
		\bibinfo {pages} {157201} (\bibinfo {year} {2017})}\BibitemShut {NoStop}%
	\bibitem [{\citenamefont {Itou}\ \emph {et~al.}(2008)\citenamefont {Itou},
		\citenamefont {Oyamada}, \citenamefont {Maegawa}, \citenamefont {Tamura},\
		and\ \citenamefont {Kato}}]{Itou2008}%
	\BibitemOpen
	\bibfield  {author} {\bibinfo {author} {\bibfnamefont {T.}~\bibnamefont
			{Itou}}, \bibinfo {author} {\bibfnamefont {A.}~\bibnamefont {Oyamada}},
		\bibinfo {author} {\bibfnamefont {S.}~\bibnamefont {Maegawa}}, \bibinfo
		{author} {\bibfnamefont {M.}~\bibnamefont {Tamura}},\ and\ \bibinfo {author}
		{\bibfnamefont {R.}~\bibnamefont {Kato}},\ }\bibfield  {title} {\bibinfo
		{title} {Quantum spin liquid in the spin-$1/2$ triangular antiferromagnet
			{$\mathrm{Et}{\mathrm{Me}}_{3}\mathrm{Sb}{[\mathrm{Pd}{(\text{dmit})}_{2}]}_{2}$}},\
	}\href {https://doi.org/10.1103/PhysRevB.77.104413} {\bibfield  {journal}
		{\bibinfo  {journal} {Phys. Rev. B}\ }\textbf {\bibinfo {volume} {77}},\
		\bibinfo {pages} {104413} (\bibinfo {year} {2008})}\BibitemShut {NoStop}%
	\bibitem [{\citenamefont {Yamashita}\ \emph {et~al.}(2008)\citenamefont
		{Yamashita}, \citenamefont {Nakazawa}, \citenamefont {Oguni}, \citenamefont
		{Oshima}, \citenamefont {Nojiri}, \citenamefont {Shimizu}, \citenamefont
		{Miyagawa},\ and\ \citenamefont {Kanoda}}]{Yamashita2008}%
	\BibitemOpen
	\bibfield  {author} {\bibinfo {author} {\bibfnamefont {S.}~\bibnamefont
			{Yamashita}}, \bibinfo {author} {\bibfnamefont {Y.}~\bibnamefont {Nakazawa}},
		\bibinfo {author} {\bibfnamefont {M.}~\bibnamefont {Oguni}}, \bibinfo
		{author} {\bibfnamefont {Y.}~\bibnamefont {Oshima}}, \bibinfo {author}
		{\bibfnamefont {H.}~\bibnamefont {Nojiri}}, \bibinfo {author} {\bibfnamefont
			{Y.}~\bibnamefont {Shimizu}}, \bibinfo {author} {\bibfnamefont
			{K.}~\bibnamefont {Miyagawa}},\ and\ \bibinfo {author} {\bibfnamefont
			{K.}~\bibnamefont {Kanoda}},\ }\bibfield  {title} {\bibinfo {title}
		{Thermodynamic properties of a spin-1/2 spin-liquid state in a k-type organic
			salt},\ }\href {https://doi.org/10.1038/nphys942} {\bibfield  {journal}
		{\bibinfo  {journal} {Nature Physics}\ }\textbf {\bibinfo {volume} {4}},\
		\bibinfo {pages} {459} (\bibinfo {year} {2008})}\BibitemShut {NoStop}%
	\bibitem [{\citenamefont {Riedl}\ \emph {et~al.}(2019)\citenamefont {Riedl},
		\citenamefont {Valent{\'i}},\ and\ \citenamefont {Winter}}]{Riedl2019}%
	\BibitemOpen
	\bibfield  {author} {\bibinfo {author} {\bibfnamefont {K.}~\bibnamefont
			{Riedl}}, \bibinfo {author} {\bibfnamefont {R.}~\bibnamefont {Valent{\'i}}},\
		and\ \bibinfo {author} {\bibfnamefont {S.~M.}\ \bibnamefont {Winter}},\
	}\bibfield  {title} {\bibinfo {title} {Critical spin liquid versus
			valence-bond glass in a triangular-lattice organic antiferromagnet},\ }\href
	{https://doi.org/10.1038/s41467-019-10604-3} {\bibfield  {journal} {\bibinfo
			{journal} {Nature Communications}\ }\textbf {\bibinfo {volume} {10}},\
		\bibinfo {pages} {2561} (\bibinfo {year} {2019})}\BibitemShut {NoStop}%
	\bibitem [{\citenamefont {Ding}\ \emph {et~al.}(2019)\citenamefont {Ding},
		\citenamefont {Manuel}, \citenamefont {Bachus}, \citenamefont {Gru\ss{}ler},
		\citenamefont {Gegenwart}, \citenamefont {Singleton}, \citenamefont
		{Johnson}, \citenamefont {Walker}, \citenamefont {Adroja}, \citenamefont
		{Hillier},\ and\ \citenamefont {Tsirlin}}]{Ding_2019_NYO}%
	\BibitemOpen
	\bibfield  {author} {\bibinfo {author} {\bibfnamefont {L.}~\bibnamefont
			{Ding}}, \bibinfo {author} {\bibfnamefont {P.}~\bibnamefont {Manuel}},
		\bibinfo {author} {\bibfnamefont {S.}~\bibnamefont {Bachus}}, \bibinfo
		{author} {\bibfnamefont {F.}~\bibnamefont {Gru\ss{}ler}}, \bibinfo {author}
		{\bibfnamefont {P.}~\bibnamefont {Gegenwart}}, \bibinfo {author}
		{\bibfnamefont {J.}~\bibnamefont {Singleton}}, \bibinfo {author}
		{\bibfnamefont {R.~D.}\ \bibnamefont {Johnson}}, \bibinfo {author}
		{\bibfnamefont {H.~C.}\ \bibnamefont {Walker}}, \bibinfo {author}
		{\bibfnamefont {D.~T.}\ \bibnamefont {Adroja}}, \bibinfo {author}
		{\bibfnamefont {A.~D.}\ \bibnamefont {Hillier}},\ and\ \bibinfo {author}
		{\bibfnamefont {A.~A.}\ \bibnamefont {Tsirlin}},\ }\bibfield  {title}
	{\bibinfo {title} {Gapless spin-liquid state in the structurally
			disorder-free triangular antiferromagnet {${\mathrm{NaYbO}}_{2}$}},\ }\href
	{https://doi.org/10.1103/PhysRevB.100.144432} {\bibfield  {journal} {\bibinfo
			{journal} {Phys. Rev. B}\ }\textbf {\bibinfo {volume} {100}},\ \bibinfo
		{pages} {144432} (\bibinfo {year} {2019})}\BibitemShut {NoStop}%
	\bibitem [{\citenamefont {Baenitz}\ \emph {et~al.}(2018)\citenamefont
		{Baenitz}, \citenamefont {Schlender}, \citenamefont {Sichelschmidt},
		\citenamefont {Onykiienko}, \citenamefont {Zangeneh}, \citenamefont
		{Ranjith}, \citenamefont {Sarkar}, \citenamefont {Hozoi}, \citenamefont
		{Walker}, \citenamefont {Orain}, \citenamefont {Yasuoka}, \citenamefont
		{van~den Brink}, \citenamefont {Klauss}, \citenamefont {Inosov},\ and\
		\citenamefont {Doert}}]{Baenitz_2018}%
	\BibitemOpen
	\bibfield  {author} {\bibinfo {author} {\bibfnamefont {M.}~\bibnamefont
			{Baenitz}}, \bibinfo {author} {\bibfnamefont {P.}~\bibnamefont {Schlender}},
		\bibinfo {author} {\bibfnamefont {J.}~\bibnamefont {Sichelschmidt}}, \bibinfo
		{author} {\bibfnamefont {Y.~A.}\ \bibnamefont {Onykiienko}}, \bibinfo
		{author} {\bibfnamefont {Z.}~\bibnamefont {Zangeneh}}, \bibinfo {author}
		{\bibfnamefont {K.~M.}\ \bibnamefont {Ranjith}}, \bibinfo {author}
		{\bibfnamefont {R.}~\bibnamefont {Sarkar}}, \bibinfo {author} {\bibfnamefont
			{L.}~\bibnamefont {Hozoi}}, \bibinfo {author} {\bibfnamefont {H.~C.}\
			\bibnamefont {Walker}}, \bibinfo {author} {\bibfnamefont {J.-C.}\
			\bibnamefont {Orain}}, \bibinfo {author} {\bibfnamefont {H.}~\bibnamefont
			{Yasuoka}}, \bibinfo {author} {\bibfnamefont {J.}~\bibnamefont {van~den
				Brink}}, \bibinfo {author} {\bibfnamefont {H.~H.}\ \bibnamefont {Klauss}},
		\bibinfo {author} {\bibfnamefont {D.~S.}\ \bibnamefont {Inosov}},\ and\
		\bibinfo {author} {\bibfnamefont {T.}~\bibnamefont {Doert}},\ }\bibfield
	{title} {\bibinfo {title} {{${\mathrm{NaYbS}}_{2}$}: A planar
			spin-$\frac{1}{2}$ triangular-lattice magnet and putative spin liquid},\
	}\href {https://doi.org/10.1103/PhysRevB.98.220409} {\bibfield  {journal}
		{\bibinfo  {journal} {Phys. Rev. B}\ }\textbf {\bibinfo {volume} {98}},\
		\bibinfo {pages} {220409} (\bibinfo {year} {2018})}\BibitemShut {NoStop}%
	\bibitem [{\citenamefont {Bordelon}\ \emph {et~al.}(2019)\citenamefont
		{Bordelon}, \citenamefont {Kenney}, \citenamefont {Liu}, \citenamefont
		{Hogan}, \citenamefont {Posthuma}, \citenamefont {Kavand}, \citenamefont
		{Lyu}, \citenamefont {Sherwin}, \citenamefont {Butch}, \citenamefont {Brown},
		\citenamefont {Graf}, \citenamefont {Balents},\ and\ \citenamefont
		{Wilson}}]{Bordelon2019}%
	\BibitemOpen
	\bibfield  {author} {\bibinfo {author} {\bibfnamefont {M.~M.}\ \bibnamefont
			{Bordelon}}, \bibinfo {author} {\bibfnamefont {E.}~\bibnamefont {Kenney}},
		\bibinfo {author} {\bibfnamefont {C.}~\bibnamefont {Liu}}, \bibinfo {author}
		{\bibfnamefont {T.}~\bibnamefont {Hogan}}, \bibinfo {author} {\bibfnamefont
			{L.}~\bibnamefont {Posthuma}}, \bibinfo {author} {\bibfnamefont
			{M.}~\bibnamefont {Kavand}}, \bibinfo {author} {\bibfnamefont
			{Y.}~\bibnamefont {Lyu}}, \bibinfo {author} {\bibfnamefont {M.}~\bibnamefont
			{Sherwin}}, \bibinfo {author} {\bibfnamefont {N.~P.}\ \bibnamefont {Butch}},
		\bibinfo {author} {\bibfnamefont {C.}~\bibnamefont {Brown}}, \bibinfo
		{author} {\bibfnamefont {M.~J.}\ \bibnamefont {Graf}}, \bibinfo {author}
		{\bibfnamefont {L.}~\bibnamefont {Balents}},\ and\ \bibinfo {author}
		{\bibfnamefont {S.~D.}\ \bibnamefont {Wilson}},\ }\bibfield  {title}
	{\bibinfo {title} {Field-tunable quantum disordered ground state in the
			triangular-lattice antiferromagnet {NaYbO$_2$}},\ }\href
	{https://doi.org/10.1038/s41567-019-0594-5} {\bibfield  {journal} {\bibinfo
			{journal} {Nature Physics}\ }\textbf {\bibinfo {volume} {15}},\ \bibinfo
		{pages} {1058} (\bibinfo {year} {2019})}\BibitemShut {NoStop}%
	\bibitem [{\citenamefont {Sarkar}\ \emph {et~al.}(2019)\citenamefont {Sarkar},
		\citenamefont {Schlender}, \citenamefont {Grinenko}, \citenamefont
		{Haeussler}, \citenamefont {Baker}, \citenamefont {Doert},\ and\
		\citenamefont {Klauss}}]{sarkar2019quantum}%
	\BibitemOpen
	\bibfield  {author} {\bibinfo {author} {\bibfnamefont {R.}~\bibnamefont
			{Sarkar}}, \bibinfo {author} {\bibfnamefont {P.}~\bibnamefont {Schlender}},
		\bibinfo {author} {\bibfnamefont {V.}~\bibnamefont {Grinenko}}, \bibinfo
		{author} {\bibfnamefont {E.}~\bibnamefont {Haeussler}}, \bibinfo {author}
		{\bibfnamefont {P.~J.}\ \bibnamefont {Baker}}, \bibinfo {author}
		{\bibfnamefont {T.}~\bibnamefont {Doert}},\ and\ \bibinfo {author}
		{\bibfnamefont {H.-H.}\ \bibnamefont {Klauss}},\ }\bibfield  {title}
	{\bibinfo {title} {Quantum spin liquid ground state in the disorder free
			triangular lattice {${\mathrm{NaYbS}}_{2}$}},\ }\href
	{https://doi.org/10.1103/PhysRevB.100.241116} {\bibfield  {journal} {\bibinfo
			{journal} {Phys. Rev. B}\ }\textbf {\bibinfo {volume} {100}},\ \bibinfo
		{pages} {241116} (\bibinfo {year} {2019})}\BibitemShut {NoStop}%
	\bibitem [{\citenamefont {Xing}\ \emph {et~al.}(2021)\citenamefont {Xing},
		\citenamefont {Sanjeewa}, \citenamefont {May},\ and\ \citenamefont
		{Sefat}}]{Xing2021_KYS}%
	\BibitemOpen
	\bibfield  {author} {\bibinfo {author} {\bibfnamefont {J.}~\bibnamefont
			{Xing}}, \bibinfo {author} {\bibfnamefont {L.~D.}\ \bibnamefont {Sanjeewa}},
		\bibinfo {author} {\bibfnamefont {A.~F.}\ \bibnamefont {May}},\ and\ \bibinfo
		{author} {\bibfnamefont {A.~S.}\ \bibnamefont {Sefat}},\ }\bibfield  {title}
	{\bibinfo {title} {Synthesis and anisotropic magnetism in quantum spin liquid
			candidates {AYbSe$_2$} {(A = K and Rb)}},\ }\href
	{https://doi.org/10.1063/5.0071161} {\bibfield  {journal} {\bibinfo
			{journal} {APL Materials}\ }\textbf {\bibinfo {volume} {9}},\ \bibinfo
		{pages} {111104} (\bibinfo {year} {2021})}\BibitemShut {NoStop}%
	\bibitem [{\citenamefont {Dai}\ \emph {et~al.}(2021)\citenamefont {Dai},
		\citenamefont {Zhang}, \citenamefont {Xie}, \citenamefont {Duan},
		\citenamefont {Gao}, \citenamefont {Zhu}, \citenamefont {Feng}, \citenamefont
		{Tao}, \citenamefont {Huang}, \citenamefont {Cao}, \citenamefont
		{Podlesnyak}, \citenamefont {Granroth}, \citenamefont {Everett},
		\citenamefont {Neuefeind}, \citenamefont {Voneshen}, \citenamefont {Wang},
		\citenamefont {Tan}, \citenamefont {Morosan}, \citenamefont {Wang},
		\citenamefont {Lin}, \citenamefont {Shu}, \citenamefont {Chen}, \citenamefont
		{Guo}, \citenamefont {Lu},\ and\ \citenamefont {Dai}}]{Dai_2021}%
	\BibitemOpen
	\bibfield  {author} {\bibinfo {author} {\bibfnamefont {P.-L.}\ \bibnamefont
			{Dai}}, \bibinfo {author} {\bibfnamefont {G.}~\bibnamefont {Zhang}}, \bibinfo
		{author} {\bibfnamefont {Y.}~\bibnamefont {Xie}}, \bibinfo {author}
		{\bibfnamefont {C.}~\bibnamefont {Duan}}, \bibinfo {author} {\bibfnamefont
			{Y.}~\bibnamefont {Gao}}, \bibinfo {author} {\bibfnamefont {Z.}~\bibnamefont
			{Zhu}}, \bibinfo {author} {\bibfnamefont {E.}~\bibnamefont {Feng}}, \bibinfo
		{author} {\bibfnamefont {Z.}~\bibnamefont {Tao}}, \bibinfo {author}
		{\bibfnamefont {C.-L.}\ \bibnamefont {Huang}}, \bibinfo {author}
		{\bibfnamefont {H.}~\bibnamefont {Cao}}, \bibinfo {author} {\bibfnamefont
			{A.}~\bibnamefont {Podlesnyak}}, \bibinfo {author} {\bibfnamefont {G.~E.}\
			\bibnamefont {Granroth}}, \bibinfo {author} {\bibfnamefont {M.~S.}\
			\bibnamefont {Everett}}, \bibinfo {author} {\bibfnamefont {J.~C.}\
			\bibnamefont {Neuefeind}}, \bibinfo {author} {\bibfnamefont {D.}~\bibnamefont
			{Voneshen}}, \bibinfo {author} {\bibfnamefont {S.}~\bibnamefont {Wang}},
		\bibinfo {author} {\bibfnamefont {G.}~\bibnamefont {Tan}}, \bibinfo {author}
		{\bibfnamefont {E.}~\bibnamefont {Morosan}}, \bibinfo {author} {\bibfnamefont
			{X.}~\bibnamefont {Wang}}, \bibinfo {author} {\bibfnamefont {H.-Q.}\
			\bibnamefont {Lin}}, \bibinfo {author} {\bibfnamefont {L.}~\bibnamefont
			{Shu}}, \bibinfo {author} {\bibfnamefont {G.}~\bibnamefont {Chen}}, \bibinfo
		{author} {\bibfnamefont {Y.}~\bibnamefont {Guo}}, \bibinfo {author}
		{\bibfnamefont {X.}~\bibnamefont {Lu}},\ and\ \bibinfo {author}
		{\bibfnamefont {P.}~\bibnamefont {Dai}},\ }\bibfield  {title} {\bibinfo
		{title} {Spinon fermi surface spin liquid in a triangular lattice
			antiferromagnet {${\mathrm{NaYbSe}}_{2}$}},\ }\href
	{https://doi.org/10.1103/PhysRevX.11.021044} {\bibfield  {journal} {\bibinfo
			{journal} {Phys. Rev. X}\ }\textbf {\bibinfo {volume} {11}},\ \bibinfo
		{pages} {021044} (\bibinfo {year} {2021})}\BibitemShut {NoStop}%
	\bibitem [{\citenamefont {Kivelson}\ \emph {et~al.}(1987)\citenamefont
		{Kivelson}, \citenamefont {Rokhsar},\ and\ \citenamefont
		{Sethna}}]{Kivelson_1987}%
	\BibitemOpen
	\bibfield  {author} {\bibinfo {author} {\bibfnamefont {S.~A.}\ \bibnamefont
			{Kivelson}}, \bibinfo {author} {\bibfnamefont {D.~S.}\ \bibnamefont
			{Rokhsar}},\ and\ \bibinfo {author} {\bibfnamefont {J.~P.}\ \bibnamefont
			{Sethna}},\ }\bibfield  {title} {\bibinfo {title} {Topology of the resonating
			valence-bond state: Solitons and high-${T}_{c}$ superconductivity},\ }\href
	{https://doi.org/10.1103/PhysRevB.35.8865} {\bibfield  {journal} {\bibinfo
			{journal} {Phys. Rev. B}\ }\textbf {\bibinfo {volume} {35}},\ \bibinfo
		{pages} {8865} (\bibinfo {year} {1987})}\BibitemShut {NoStop}%
	\bibitem [{\citenamefont {Anderson}\ \emph {et~al.}(1987)\citenamefont
		{Anderson}, \citenamefont {Baskaran}, \citenamefont {Zou},\ and\
		\citenamefont {Hsu}}]{Anderson_1987}%
	\BibitemOpen
	\bibfield  {author} {\bibinfo {author} {\bibfnamefont {P.~W.}\ \bibnamefont
			{Anderson}}, \bibinfo {author} {\bibfnamefont {G.}~\bibnamefont {Baskaran}},
		\bibinfo {author} {\bibfnamefont {Z.}~\bibnamefont {Zou}},\ and\ \bibinfo
		{author} {\bibfnamefont {T.}~\bibnamefont {Hsu}},\ }\bibfield  {title}
	{\bibinfo {title} {Resonating--valence-bond theory of phase transitions and
			superconductivity in ${\mathrm{la}}_{2}$${\mathrm{cuo}}_{4}$-based
			compounds},\ }\href {https://doi.org/10.1103/PhysRevLett.58.2790} {\bibfield
		{journal} {\bibinfo  {journal} {Phys. Rev. Lett.}\ }\textbf {\bibinfo
			{volume} {58}},\ \bibinfo {pages} {2790} (\bibinfo {year}
		{1987})}\BibitemShut {NoStop}%
	\bibitem [{\citenamefont {Kimchi}\ \emph {et~al.}(2018)\citenamefont {Kimchi},
		\citenamefont {Nahum},\ and\ \citenamefont {Senthil}}]{Kimchi_2018}%
	\BibitemOpen
	\bibfield  {author} {\bibinfo {author} {\bibfnamefont {I.}~\bibnamefont
			{Kimchi}}, \bibinfo {author} {\bibfnamefont {A.}~\bibnamefont {Nahum}},\ and\
		\bibinfo {author} {\bibfnamefont {T.}~\bibnamefont {Senthil}},\ }\bibfield
	{title} {\bibinfo {title} {Valence bonds in random quantum magnets: Theory
			and application to {${\mathrm{YbMgGaO}}_{4}$}},\ }\href
	{https://doi.org/10.1103/PhysRevX.8.031028} {\bibfield  {journal} {\bibinfo
			{journal} {Phys. Rev. X}\ }\textbf {\bibinfo {volume} {8}},\ \bibinfo {pages}
		{031028} (\bibinfo {year} {2018})}\BibitemShut {NoStop}%
	\bibitem [{\citenamefont {Nishimori}\ and\ \citenamefont
		{Miyashita}(1986)}]{Nishimori_1986}%
	\BibitemOpen
	\bibfield  {author} {\bibinfo {author} {\bibfnamefont {H.}~\bibnamefont
			{Nishimori}}\ and\ \bibinfo {author} {\bibfnamefont {S.}~\bibnamefont
			{Miyashita}},\ }\bibfield  {title} {\bibinfo {title} {Magnetization process
			of the spin-1/2 antiferromagnetic ising-like heisenberg model on the
			triangular lattice},\ }\href {https://doi.org/10.1143/JPSJ.55.4448}
	{\bibfield  {journal} {\bibinfo  {journal} {Journal of the Physical Society
				of Japan}\ }\textbf {\bibinfo {volume} {55}},\ \bibinfo {pages} {4448}
		(\bibinfo {year} {1986})}\BibitemShut {NoStop}%
	\bibitem [{\citenamefont {Chubukov}\ and\ \citenamefont
		{Golosov}(1991)}]{Chubukov_1991}%
	\BibitemOpen
	\bibfield  {author} {\bibinfo {author} {\bibfnamefont {A.~V.}\ \bibnamefont
			{Chubukov}}\ and\ \bibinfo {author} {\bibfnamefont {D.~I.}\ \bibnamefont
			{Golosov}},\ }\bibfield  {title} {\bibinfo {title} {Quantum theory of an
			antiferromagnet on a triangular lattice in a magnetic field},\ }\href
	{https://doi.org/10.1088/0953-8984/3/1/005} {\bibfield  {journal} {\bibinfo
			{journal} {Journal of Physics: Condensed Matter}\ }\textbf {\bibinfo {volume}
			{3}},\ \bibinfo {pages} {69} (\bibinfo {year} {1991})}\BibitemShut {NoStop}%
	\bibitem [{\citenamefont {Alicea}\ \emph {et~al.}(2009)\citenamefont {Alicea},
		\citenamefont {Chubukov},\ and\ \citenamefont {Starykh}}]{Alicea_2009}%
	\BibitemOpen
	\bibfield  {author} {\bibinfo {author} {\bibfnamefont {J.}~\bibnamefont
			{Alicea}}, \bibinfo {author} {\bibfnamefont {A.~V.}\ \bibnamefont
			{Chubukov}},\ and\ \bibinfo {author} {\bibfnamefont {O.~A.}\ \bibnamefont
			{Starykh}},\ }\bibfield  {title} {\bibinfo {title} {Quantum stabilization of
			the $1/3$-magnetization plateau in ${\mathrm{cs}}_{2}{\mathrm{cubr}}_{4}$},\
	}\href {https://doi.org/10.1103/PhysRevLett.102.137201} {\bibfield  {journal}
		{\bibinfo  {journal} {Phys. Rev. Lett.}\ }\textbf {\bibinfo {volume} {102}},\
		\bibinfo {pages} {137201} (\bibinfo {year} {2009})}\BibitemShut {NoStop}%
	\bibitem [{\citenamefont {Kamiya}\ \emph {et~al.}(2018)\citenamefont {Kamiya},
		\citenamefont {Ge}, \citenamefont {Hong}, \citenamefont {Qiu}, \citenamefont
		{Quintero-Castro}, \citenamefont {Lu}, \citenamefont {Cao}, \citenamefont
		{Matsuda}, \citenamefont {Choi}, \citenamefont {Batista}, \citenamefont
		{Mourigal}, \citenamefont {Zhou},\ and\ \citenamefont {Ma}}]{Kamiya18}%
	\BibitemOpen
	\bibfield  {author} {\bibinfo {author} {\bibfnamefont {Y.}~\bibnamefont
			{Kamiya}}, \bibinfo {author} {\bibfnamefont {L.}~\bibnamefont {Ge}}, \bibinfo
		{author} {\bibfnamefont {T.}~\bibnamefont {Hong}}, \bibinfo {author}
		{\bibfnamefont {Y.}~\bibnamefont {Qiu}}, \bibinfo {author} {\bibfnamefont
			{D.~L.}\ \bibnamefont {Quintero-Castro}}, \bibinfo {author} {\bibfnamefont
			{Z.}~\bibnamefont {Lu}}, \bibinfo {author} {\bibfnamefont {H.~B.}\
			\bibnamefont {Cao}}, \bibinfo {author} {\bibfnamefont {M.}~\bibnamefont
			{Matsuda}}, \bibinfo {author} {\bibfnamefont {E.~S.}\ \bibnamefont {Choi}},
		\bibinfo {author} {\bibfnamefont {C.~D.}\ \bibnamefont {Batista}}, \bibinfo
		{author} {\bibfnamefont {M.}~\bibnamefont {Mourigal}}, \bibinfo {author}
		{\bibfnamefont {H.~D.}\ \bibnamefont {Zhou}},\ and\ \bibinfo {author}
		{\bibfnamefont {J.}~\bibnamefont {Ma}},\ }\bibfield  {title} {\bibinfo
		{title} {The nature of spin excitations in the one-third magnetization
			plateau phase of {Ba$_3$CoSb$_2$O$_9$}},\ }\href
	{https://doi.org/10.1038/s41467-018-04914-1} {\bibfield  {journal} {\bibinfo
			{journal} {Nature Communications}\ }\textbf {\bibinfo {volume} {9}},\
		\bibinfo {pages} {2666} (\bibinfo {year} {2018})}\BibitemShut {NoStop}%
	\bibitem [{\citenamefont {Coldea}\ \emph {et~al.}(2002)\citenamefont {Coldea},
		\citenamefont {Tennant}, \citenamefont {Habicht}, \citenamefont {Smeibidl},
		\citenamefont {Wolters},\ and\ \citenamefont {Tylczynski}}]{Coldea_2002}%
	\BibitemOpen
	\bibfield  {author} {\bibinfo {author} {\bibfnamefont {R.}~\bibnamefont
			{Coldea}}, \bibinfo {author} {\bibfnamefont {D.~A.}\ \bibnamefont {Tennant}},
		\bibinfo {author} {\bibfnamefont {K.}~\bibnamefont {Habicht}}, \bibinfo
		{author} {\bibfnamefont {P.}~\bibnamefont {Smeibidl}}, \bibinfo {author}
		{\bibfnamefont {C.}~\bibnamefont {Wolters}},\ and\ \bibinfo {author}
		{\bibfnamefont {Z.}~\bibnamefont {Tylczynski}},\ }\bibfield  {title}
	{\bibinfo {title} {Direct measurement of the spin hamiltonian and observation
			of condensation of magnons in the 2d frustrated quantum magnet
			{${\mathrm{Cs}}_{2}{\mathrm{CuCl}}_{4}$}},\ }\href
	{https://doi.org/10.1103/PhysRevLett.88.137203} {\bibfield  {journal}
		{\bibinfo  {journal} {Phys. Rev. Lett.}\ }\textbf {\bibinfo {volume} {88}},\
		\bibinfo {pages} {137203} (\bibinfo {year} {2002})}\BibitemShut {NoStop}%
	\bibitem [{\citenamefont {Liu}\ \emph {et~al.}(2018)\citenamefont {Liu},
		\citenamefont {Zhang}, \citenamefont {Ji}, \citenamefont {Liu}, \citenamefont
		{Li}, \citenamefont {Wang}, \citenamefont {Lei}, \citenamefont {Chen},\ and\
		\citenamefont {Zhang}}]{Liu_2018_Chalcogenides}%
	\BibitemOpen
	\bibfield  {author} {\bibinfo {author} {\bibfnamefont {W.}~\bibnamefont
			{Liu}}, \bibinfo {author} {\bibfnamefont {Z.}~\bibnamefont {Zhang}}, \bibinfo
		{author} {\bibfnamefont {J.}~\bibnamefont {Ji}}, \bibinfo {author}
		{\bibfnamefont {Y.}~\bibnamefont {Liu}}, \bibinfo {author} {\bibfnamefont
			{J.}~\bibnamefont {Li}}, \bibinfo {author} {\bibfnamefont {X.}~\bibnamefont
			{Wang}}, \bibinfo {author} {\bibfnamefont {H.}~\bibnamefont {Lei}}, \bibinfo
		{author} {\bibfnamefont {G.}~\bibnamefont {Chen}},\ and\ \bibinfo {author}
		{\bibfnamefont {Q.}~\bibnamefont {Zhang}},\ }\bibfield  {title} {\bibinfo
		{title} {Rare-earth chalcogenides: A large family of triangular lattice spin
			liquid candidates},\ }\href {https://doi.org/10.1088/0256-307x/35/11/117501}
	{\bibfield  {journal} {\bibinfo  {journal} {Chinese Physics Letters}\
		}\textbf {\bibinfo {volume} {35}},\ \bibinfo {pages} {117501} (\bibinfo
		{year} {2018})}\BibitemShut {NoStop}%
	\bibitem [{\citenamefont {Ranjith}\ \emph {et~al.}(2019)\citenamefont
		{Ranjith}, \citenamefont {Luther}, \citenamefont {Reimann}, \citenamefont
		{Schmidt}, \citenamefont {Schlender}, \citenamefont {Sichelschmidt},
		\citenamefont {Yasuoka}, \citenamefont {Strydom}, \citenamefont {Skourski},
		\citenamefont {Wosnitza}, \citenamefont {K\"uhne}, \citenamefont {Doert},\
		and\ \citenamefont {Baenitz}}]{Ranjith2019_2}%
	\BibitemOpen
	\bibfield  {author} {\bibinfo {author} {\bibfnamefont {K.~M.}\ \bibnamefont
			{Ranjith}}, \bibinfo {author} {\bibfnamefont {S.}~\bibnamefont {Luther}},
		\bibinfo {author} {\bibfnamefont {T.}~\bibnamefont {Reimann}}, \bibinfo
		{author} {\bibfnamefont {B.}~\bibnamefont {Schmidt}}, \bibinfo {author}
		{\bibfnamefont {P.}~\bibnamefont {Schlender}}, \bibinfo {author}
		{\bibfnamefont {J.}~\bibnamefont {Sichelschmidt}}, \bibinfo {author}
		{\bibfnamefont {H.}~\bibnamefont {Yasuoka}}, \bibinfo {author} {\bibfnamefont
			{A.~M.}\ \bibnamefont {Strydom}}, \bibinfo {author} {\bibfnamefont
			{Y.}~\bibnamefont {Skourski}}, \bibinfo {author} {\bibfnamefont
			{J.}~\bibnamefont {Wosnitza}}, \bibinfo {author} {\bibfnamefont
			{H.}~\bibnamefont {K\"uhne}}, \bibinfo {author} {\bibfnamefont
			{T.}~\bibnamefont {Doert}},\ and\ \bibinfo {author} {\bibfnamefont
			{M.}~\bibnamefont {Baenitz}},\ }\bibfield  {title} {\bibinfo {title}
		{Anisotropic field-induced ordering in the triangular-lattice quantum spin
			liquid {${\mathrm{NaYbSe}}_{2}$}},\ }\href
	{https://doi.org/10.1103/PhysRevB.100.224417} {\bibfield  {journal} {\bibinfo
			{journal} {Phys. Rev. B}\ }\textbf {\bibinfo {volume} {100}},\ \bibinfo
		{pages} {224417} (\bibinfo {year} {2019})}\BibitemShut {NoStop}%
	\bibitem [{\citenamefont {Scheie}\ \emph {et~al.}(2021)\citenamefont {Scheie},
		\citenamefont {Ghioldi}, \citenamefont {Xing}, \citenamefont {Paddison},
		\citenamefont {Sherman}, \citenamefont {Dupont}, \citenamefont {Abernathy},
		\citenamefont {Pajerowski}, \citenamefont {Zhang}, \citenamefont {Manuel}
		\emph {et~al.}}]{scheie2021_KYS}%
	\BibitemOpen
	\bibfield  {author} {\bibinfo {author} {\bibfnamefont {A.}~\bibnamefont
			{Scheie}}, \bibinfo {author} {\bibfnamefont {E.}~\bibnamefont {Ghioldi}},
		\bibinfo {author} {\bibfnamefont {J.}~\bibnamefont {Xing}}, \bibinfo {author}
		{\bibfnamefont {J.}~\bibnamefont {Paddison}}, \bibinfo {author}
		{\bibfnamefont {N.}~\bibnamefont {Sherman}}, \bibinfo {author} {\bibfnamefont
			{M.}~\bibnamefont {Dupont}}, \bibinfo {author} {\bibfnamefont
			{D.}~\bibnamefont {Abernathy}}, \bibinfo {author} {\bibfnamefont
			{D.}~\bibnamefont {Pajerowski}}, \bibinfo {author} {\bibfnamefont {S.-S.}\
			\bibnamefont {Zhang}}, \bibinfo {author} {\bibfnamefont {L.}~\bibnamefont
			{Manuel}}, \emph {et~al.},\ }\bibfield  {title} {\bibinfo {title} {Witnessing
			quantum criticality and entanglement in the triangular antiferromagnet kybse
			$ \_2$},\ }\href {https://arxiv.org/abs/2109.11527} {\bibfield  {journal}
		{\bibinfo  {journal} {arXiv preprint arXiv:2109.11527}\ } (\bibinfo {year}
		{2021})}\BibitemShut {NoStop}%
	\bibitem [{\citenamefont {Zhang}\ \emph {et~al.}(2022)\citenamefont {Zhang},
		\citenamefont {Li}, \citenamefont {Xie}, \citenamefont {Zhuo}, \citenamefont
		{Adroja}, \citenamefont {Baker}, \citenamefont {Perring}, \citenamefont
		{Zhang}, \citenamefont {Jin}, \citenamefont {Ji}, \citenamefont {Wang},
		\citenamefont {Ma},\ and\ \citenamefont {Zhang}}]{Zhang_2022_NYS}%
	\BibitemOpen
	\bibfield  {author} {\bibinfo {author} {\bibfnamefont {Z.}~\bibnamefont
			{Zhang}}, \bibinfo {author} {\bibfnamefont {J.}~\bibnamefont {Li}}, \bibinfo
		{author} {\bibfnamefont {M.}~\bibnamefont {Xie}}, \bibinfo {author}
		{\bibfnamefont {W.}~\bibnamefont {Zhuo}}, \bibinfo {author} {\bibfnamefont
			{D.~T.}\ \bibnamefont {Adroja}}, \bibinfo {author} {\bibfnamefont {P.~J.}\
			\bibnamefont {Baker}}, \bibinfo {author} {\bibfnamefont {T.~G.}\ \bibnamefont
			{Perring}}, \bibinfo {author} {\bibfnamefont {A.}~\bibnamefont {Zhang}},
		\bibinfo {author} {\bibfnamefont {F.}~\bibnamefont {Jin}}, \bibinfo {author}
		{\bibfnamefont {J.}~\bibnamefont {Ji}}, \bibinfo {author} {\bibfnamefont
			{X.}~\bibnamefont {Wang}}, \bibinfo {author} {\bibfnamefont {J.}~\bibnamefont
			{Ma}},\ and\ \bibinfo {author} {\bibfnamefont {Q.}~\bibnamefont {Zhang}},\
	}\bibfield  {title} {\bibinfo {title} {Low-energy spin dynamics of the
			quantum spin liquid candidate {${\mathrm{NaYbSe}}_{2}$}},\ }\href
	{https://doi.org/10.1103/PhysRevB.106.085115} {\bibfield  {journal} {\bibinfo
			{journal} {Phys. Rev. B}\ }\textbf {\bibinfo {volume} {106}},\ \bibinfo
		{pages} {085115} (\bibinfo {year} {2022})}\BibitemShut {NoStop}%
	\bibitem [{\citenamefont {Ehlers}\ \emph {et~al.}(2011)\citenamefont {Ehlers},
		\citenamefont {Podlesnyak}, \citenamefont {Niedziela}, \citenamefont
		{Iverson},\ and\ \citenamefont {Sokol}}]{CNCS}%
	\BibitemOpen
	\bibfield  {author} {\bibinfo {author} {\bibfnamefont {G.}~\bibnamefont
			{Ehlers}}, \bibinfo {author} {\bibfnamefont {A.~A.}\ \bibnamefont
			{Podlesnyak}}, \bibinfo {author} {\bibfnamefont {J.~L.}\ \bibnamefont
			{Niedziela}}, \bibinfo {author} {\bibfnamefont {E.~B.}\ \bibnamefont
			{Iverson}},\ and\ \bibinfo {author} {\bibfnamefont {P.~E.}\ \bibnamefont
			{Sokol}},\ }\bibfield  {title} {\bibinfo {title} {The new cold neutron
			chopper spectrometer at the spallation neutron source: Design and
			performance},\ }\href {https://doi.org/10.1063/1.3626935} {\bibfield
		{journal} {\bibinfo  {journal} {Review of Scientific Instruments}\ }\textbf
		{\bibinfo {volume} {82}},\ \bibinfo {pages} {085108} (\bibinfo {year}
		{2011})}\BibitemShut {NoStop}%
	\bibitem [{\citenamefont {Mason}\ \emph {et~al.}(2006)\citenamefont {Mason},
		\citenamefont {Abernathy}, \citenamefont {Anderson}, \citenamefont {Ankner},
		\citenamefont {Egami}, \citenamefont {Ehlers}, \citenamefont {Ekkebus},
		\citenamefont {Granroth}, \citenamefont {Hagen}, \citenamefont {Herwig},
		\citenamefont {Hodges}, \citenamefont {Hoffmann}, \citenamefont {Horak},
		\citenamefont {Horton}, \citenamefont {Klose}, \citenamefont {Larese},
		\citenamefont {Mesecar}, \citenamefont {Myles}, \citenamefont {Neuefeind},
		\citenamefont {Ohl}, \citenamefont {Tulk}, \citenamefont {Wang},\ and\
		\citenamefont {Zhao}}]{mason2006spallation}%
	\BibitemOpen
	\bibfield  {author} {\bibinfo {author} {\bibfnamefont {T.~E.}\ \bibnamefont
			{Mason}}, \bibinfo {author} {\bibfnamefont {D.}~\bibnamefont {Abernathy}},
		\bibinfo {author} {\bibfnamefont {I.}~\bibnamefont {Anderson}}, \bibinfo
		{author} {\bibfnamefont {J.}~\bibnamefont {Ankner}}, \bibinfo {author}
		{\bibfnamefont {T.}~\bibnamefont {Egami}}, \bibinfo {author} {\bibfnamefont
			{G.}~\bibnamefont {Ehlers}}, \bibinfo {author} {\bibfnamefont
			{A.}~\bibnamefont {Ekkebus}}, \bibinfo {author} {\bibfnamefont
			{G.}~\bibnamefont {Granroth}}, \bibinfo {author} {\bibfnamefont
			{M.}~\bibnamefont {Hagen}}, \bibinfo {author} {\bibfnamefont
			{K.}~\bibnamefont {Herwig}}, \bibinfo {author} {\bibfnamefont
			{J.}~\bibnamefont {Hodges}}, \bibinfo {author} {\bibfnamefont
			{C.}~\bibnamefont {Hoffmann}}, \bibinfo {author} {\bibfnamefont
			{C.}~\bibnamefont {Horak}}, \bibinfo {author} {\bibfnamefont
			{L.}~\bibnamefont {Horton}}, \bibinfo {author} {\bibfnamefont
			{F.}~\bibnamefont {Klose}}, \bibinfo {author} {\bibfnamefont
			{J.}~\bibnamefont {Larese}}, \bibinfo {author} {\bibfnamefont
			{A.}~\bibnamefont {Mesecar}}, \bibinfo {author} {\bibfnamefont
			{D.}~\bibnamefont {Myles}}, \bibinfo {author} {\bibfnamefont
			{J.}~\bibnamefont {Neuefeind}}, \bibinfo {author} {\bibfnamefont
			{M.}~\bibnamefont {Ohl}}, \bibinfo {author} {\bibfnamefont {C.}~\bibnamefont
			{Tulk}}, \bibinfo {author} {\bibfnamefont {X.-L.}\ \bibnamefont {Wang}},\
		and\ \bibinfo {author} {\bibfnamefont {J.}~\bibnamefont {Zhao}},\ }\bibfield
	{title} {\bibinfo {title} {The spallation neutron source in oak ridge: A
			powerful tool for materials research},\ }\href
	{https://doi.org/10.1016/j.physb.2006.05.281} {\bibfield  {journal} {\bibinfo
			{journal} {Physica B: Condensed Matter}\ }\textbf {\bibinfo {volume}
			{385}},\ \bibinfo {pages} {955} (\bibinfo {year} {2006})}\BibitemShut
	{NoStop}%
	\bibitem [{\citenamefont {Xie}\ \emph {et~al.}(2021)\citenamefont {Xie},
		\citenamefont {Xing}, \citenamefont {Nikitin}, \citenamefont {Nishimoto},
		\citenamefont {Brando}, \citenamefont {Khanenko}, \citenamefont
		{Sichelschmidt}, \citenamefont {Sanjeewa}, \citenamefont {Sefat},\ and\
		\citenamefont {Podlesnyak}}]{xie2021field}%
	\BibitemOpen
	\bibfield  {author} {\bibinfo {author} {\bibfnamefont {T.}~\bibnamefont
			{Xie}}, \bibinfo {author} {\bibfnamefont {J.}~\bibnamefont {Xing}}, \bibinfo
		{author} {\bibfnamefont {S.}~\bibnamefont {Nikitin}}, \bibinfo {author}
		{\bibfnamefont {S.}~\bibnamefont {Nishimoto}}, \bibinfo {author}
		{\bibfnamefont {M.}~\bibnamefont {Brando}}, \bibinfo {author} {\bibfnamefont
			{P.}~\bibnamefont {Khanenko}}, \bibinfo {author} {\bibfnamefont
			{J.}~\bibnamefont {Sichelschmidt}}, \bibinfo {author} {\bibfnamefont
			{L.}~\bibnamefont {Sanjeewa}}, \bibinfo {author} {\bibfnamefont {A.~S.}\
			\bibnamefont {Sefat}},\ and\ \bibinfo {author} {\bibfnamefont
			{A.}~\bibnamefont {Podlesnyak}},\ }\bibfield  {title} {\bibinfo {title}
		{Field-induced spin excitations in the spin-1/2 triangular-lattice
			antiferromagnet {CsYbSe$_2$}},\ }\href {https://arxiv.org/abs/2106.12451}
	{\bibfield  {journal} {\bibinfo  {journal} {arXiv preprint arXiv:2106.12451}\
		} (\bibinfo {year} {2021})}\BibitemShut {NoStop}%
	\bibitem [{\citenamefont {Xie}\ \emph {et~al.}(2022)\citenamefont {Xie},
		\citenamefont {Eberharter}, \citenamefont {Xing}, \citenamefont {Nishimoto},
		\citenamefont {Brando}, \citenamefont {Khanenko}, \citenamefont
		{Sichelschmidt}, \citenamefont {Turrini}, \citenamefont {Mazzone},
		\citenamefont {Naumov} \emph {et~al.}}]{xie2022complete}%
	\BibitemOpen
	\bibfield  {author} {\bibinfo {author} {\bibfnamefont {T.}~\bibnamefont
			{Xie}}, \bibinfo {author} {\bibfnamefont {A.}~\bibnamefont {Eberharter}},
		\bibinfo {author} {\bibfnamefont {J.}~\bibnamefont {Xing}}, \bibinfo {author}
		{\bibfnamefont {S.}~\bibnamefont {Nishimoto}}, \bibinfo {author}
		{\bibfnamefont {M.}~\bibnamefont {Brando}}, \bibinfo {author} {\bibfnamefont
			{P.}~\bibnamefont {Khanenko}}, \bibinfo {author} {\bibfnamefont
			{J.}~\bibnamefont {Sichelschmidt}}, \bibinfo {author} {\bibfnamefont
			{A.}~\bibnamefont {Turrini}}, \bibinfo {author} {\bibfnamefont
			{D.}~\bibnamefont {Mazzone}}, \bibinfo {author} {\bibfnamefont
			{P.}~\bibnamefont {Naumov}}, \emph {et~al.},\ }\bibfield  {title} {\bibinfo
		{title} {Complete field-induced spectral response of the spin-1/2
			triangular-lattice antiferromagnet {CsYbSe$_2$}},\ }\href
	{https://arxiv.org/abs/2210.04928} {\bibfield  {journal} {\bibinfo  {journal}
			{arXiv preprint arXiv:2210.04928}\ } (\bibinfo {year} {2022})}\BibitemShut
	{NoStop}%
	\bibitem [{\citenamefont {Scheie}(2022)}]{scheie2022quantifying}%
	\BibitemOpen
	\bibfield  {author} {\bibinfo {author} {\bibfnamefont {A.}~\bibnamefont
			{Scheie}},\ }\bibfield  {title} {\bibinfo {title} {{Quantifying uncertainties
				in crystal electric field Hamiltonian fits to neutron data}},\ }\href
	{https://doi.org/10.21468/SciPostPhysCore.5.1.018} {\bibfield  {journal}
		{\bibinfo  {journal} {SciPost Phys. Core}\ }\textbf {\bibinfo {volume} {5}},\
		\bibinfo {pages} {18} (\bibinfo {year} {2022})}\BibitemShut {NoStop}%
	\bibitem [{\citenamefont {Press}\ \emph {et~al.}(2007)\citenamefont {Press},
		\citenamefont {Teukolsky}, \citenamefont {Vetterling},\ and\ \citenamefont
		{Flannery}}]{NumericalRecipes}%
	\BibitemOpen
	\bibfield  {author} {\bibinfo {author} {\bibfnamefont {W.~H.}\ \bibnamefont
			{Press}}, \bibinfo {author} {\bibfnamefont {S.~A.}\ \bibnamefont
			{Teukolsky}}, \bibinfo {author} {\bibfnamefont {W.~T.}\ \bibnamefont
			{Vetterling}},\ and\ \bibinfo {author} {\bibfnamefont {B.~P.}\ \bibnamefont
			{Flannery}},\ }\href@noop {} {\emph {\bibinfo {title} {Numerical recipes 3rd
				edition: The art of scientific computing}}}\ (\bibinfo  {publisher}
	{Cambridge university press},\ \bibinfo {year} {2007})\BibitemShut {NoStop}%
	\bibitem [{\citenamefont {Sala}\ \emph {et~al.}(2021)\citenamefont {Sala},
		\citenamefont {Stone}, \citenamefont {Rai}, \citenamefont {May},
		\citenamefont {Laurell}, \citenamefont {Garlea}, \citenamefont {Butch},
		\citenamefont {Lumsden}, \citenamefont {Ehlers}, \citenamefont {Pokharel}
		\emph {et~al.}}]{sala2021van}%
	\BibitemOpen
	\bibfield  {author} {\bibinfo {author} {\bibfnamefont {G.}~\bibnamefont
			{Sala}}, \bibinfo {author} {\bibfnamefont {M.~B.}\ \bibnamefont {Stone}},
		\bibinfo {author} {\bibfnamefont {B.~K.}\ \bibnamefont {Rai}}, \bibinfo
		{author} {\bibfnamefont {A.~F.}\ \bibnamefont {May}}, \bibinfo {author}
		{\bibfnamefont {P.}~\bibnamefont {Laurell}}, \bibinfo {author} {\bibfnamefont
			{V.~O.}\ \bibnamefont {Garlea}}, \bibinfo {author} {\bibfnamefont {N.~P.}\
			\bibnamefont {Butch}}, \bibinfo {author} {\bibfnamefont {M.~D.}\ \bibnamefont
			{Lumsden}}, \bibinfo {author} {\bibfnamefont {G.}~\bibnamefont {Ehlers}},
		\bibinfo {author} {\bibfnamefont {G.}~\bibnamefont {Pokharel}}, \emph
		{et~al.},\ }\bibfield  {title} {\bibinfo {title} {Van hove singularity in the
			magnon spectrum of the antiferromagnetic quantum honeycomb lattice},\ }\href
	{https://doi.org/10.1038/s41467-020-20335-5} {\bibfield  {journal} {\bibinfo
			{journal} {Nature communications}\ }\textbf {\bibinfo {volume} {12}},\
		\bibinfo {pages} {1} (\bibinfo {year} {2021})}\BibitemShut {NoStop}%
	\bibitem [{\citenamefont {Bernu}\ and\ \citenamefont
		{Misguich}(2001)}]{Bernu_2001}%
	\BibitemOpen
	\bibfield  {author} {\bibinfo {author} {\bibfnamefont {B.}~\bibnamefont
			{Bernu}}\ and\ \bibinfo {author} {\bibfnamefont {G.}~\bibnamefont
			{Misguich}},\ }\bibfield  {title} {\bibinfo {title} {Specific heat and
			high-temperature series of lattice models: Interpolation scheme and examples
			on quantum spin systems in one and two dimensions},\ }\href
	{https://doi.org/10.1103/PhysRevB.63.134409} {\bibfield  {journal} {\bibinfo
			{journal} {Phys. Rev. B}\ }\textbf {\bibinfo {volume} {63}},\ \bibinfo
		{pages} {134409} (\bibinfo {year} {2001})}\BibitemShut {NoStop}%
	\bibitem [{\citenamefont {Schmidt}\ \emph {et~al.}(2017)\citenamefont
		{Schmidt}, \citenamefont {Hauser}, \citenamefont {Lohmann},\ and\
		\citenamefont {Richter}}]{Schmidt_2017}%
	\BibitemOpen
	\bibfield  {author} {\bibinfo {author} {\bibfnamefont {H.-J.}\ \bibnamefont
			{Schmidt}}, \bibinfo {author} {\bibfnamefont {A.}~\bibnamefont {Hauser}},
		\bibinfo {author} {\bibfnamefont {A.}~\bibnamefont {Lohmann}},\ and\ \bibinfo
		{author} {\bibfnamefont {J.}~\bibnamefont {Richter}},\ }\bibfield  {title}
	{\bibinfo {title} {Interpolation between low and high temperatures of the
			specific heat for spin systems},\ }\href
	{https://doi.org/10.1103/PhysRevE.95.042110} {\bibfield  {journal} {\bibinfo
			{journal} {Phys. Rev. E}\ }\textbf {\bibinfo {volume} {95}},\ \bibinfo
		{pages} {042110} (\bibinfo {year} {2017})}\BibitemShut {NoStop}%
	\bibitem [{\citenamefont {Bernu}\ \emph {et~al.}(2020)\citenamefont {Bernu},
		\citenamefont {Pierre}, \citenamefont {Essafi},\ and\ \citenamefont
		{Messio}}]{Bernu_2020}%
	\BibitemOpen
	\bibfield  {author} {\bibinfo {author} {\bibfnamefont {B.}~\bibnamefont
			{Bernu}}, \bibinfo {author} {\bibfnamefont {L.}~\bibnamefont {Pierre}},
		\bibinfo {author} {\bibfnamefont {K.}~\bibnamefont {Essafi}},\ and\ \bibinfo
		{author} {\bibfnamefont {L.}~\bibnamefont {Messio}},\ }\bibfield  {title}
	{\bibinfo {title} {Effect of perturbations on the kagome $s=\frac{1}{2}$
			antiferromagnet at all temperatures},\ }\href
	{https://doi.org/10.1103/PhysRevB.101.140403} {\bibfield  {journal} {\bibinfo
			{journal} {Phys. Rev. B}\ }\textbf {\bibinfo {volume} {101}},\ \bibinfo
		{pages} {140403} (\bibinfo {year} {2020})}\BibitemShut {NoStop}%
	\bibitem [{\citenamefont {Gonzalez}\ \emph {et~al.}(2022)\citenamefont
		{Gonzalez}, \citenamefont {Bernu}, \citenamefont {Pierre},\ and\
		\citenamefont {Messio}}]{gonzalez2021ground}%
	\BibitemOpen
	\bibfield  {author} {\bibinfo {author} {\bibfnamefont {M.~G.}\ \bibnamefont
			{Gonzalez}}, \bibinfo {author} {\bibfnamefont {B.}~\bibnamefont {Bernu}},
		\bibinfo {author} {\bibfnamefont {L.}~\bibnamefont {Pierre}},\ and\ \bibinfo
		{author} {\bibfnamefont {L.}~\bibnamefont {Messio}},\ }\bibfield  {title}
	{\bibinfo {title} {{Ground-state and thermodynamic properties of the
				spin-$\frac{1}{2}$ Heisenberg model on the anisotropic triangular lattice}},\
	}\href {https://doi.org/10.21468/SciPostPhys.12.3.112} {\bibfield  {journal}
		{\bibinfo  {journal} {SciPost Phys.}\ }\textbf {\bibinfo {volume} {12}},\
		\bibinfo {pages} {112} (\bibinfo {year} {2022})}\BibitemShut {NoStop}%
	\bibitem [{\citenamefont {Mermin}\ and\ \citenamefont
		{Wagner}(1966)}]{MerminWagner}%
	\BibitemOpen
	\bibfield  {author} {\bibinfo {author} {\bibfnamefont {N.~D.}\ \bibnamefont
			{Mermin}}\ and\ \bibinfo {author} {\bibfnamefont {H.}~\bibnamefont
			{Wagner}},\ }\bibfield  {title} {\bibinfo {title} {Absence of ferromagnetism
			or antiferromagnetism in one- or two-dimensional isotropic heisenberg
			models},\ }\href {https://doi.org/10.1103/PhysRevLett.17.1133} {\bibfield
		{journal} {\bibinfo  {journal} {Phys. Rev. Lett.}\ }\textbf {\bibinfo
			{volume} {17}},\ \bibinfo {pages} {1133} (\bibinfo {year}
		{1966})}\BibitemShut {NoStop}%
	\bibitem [{\citenamefont {Chen}\ \emph {et~al.}(2019)\citenamefont {Chen},
		\citenamefont {Qu}, \citenamefont {Li}, \citenamefont {Chen}, \citenamefont
		{Gong}, \citenamefont {von Delft}, \citenamefont {Weichselbaum},\ and\
		\citenamefont {Li}}]{Chen_2019_twotemperature}%
	\BibitemOpen
	\bibfield  {author} {\bibinfo {author} {\bibfnamefont {L.}~\bibnamefont
			{Chen}}, \bibinfo {author} {\bibfnamefont {D.-W.}\ \bibnamefont {Qu}},
		\bibinfo {author} {\bibfnamefont {H.}~\bibnamefont {Li}}, \bibinfo {author}
		{\bibfnamefont {B.-B.}\ \bibnamefont {Chen}}, \bibinfo {author}
		{\bibfnamefont {S.-S.}\ \bibnamefont {Gong}}, \bibinfo {author}
		{\bibfnamefont {J.}~\bibnamefont {von Delft}}, \bibinfo {author}
		{\bibfnamefont {A.}~\bibnamefont {Weichselbaum}},\ and\ \bibinfo {author}
		{\bibfnamefont {W.}~\bibnamefont {Li}},\ }\bibfield  {title} {\bibinfo
		{title} {Two-temperature scales in the triangular-lattice heisenberg
			antiferromagnet},\ }\href {https://doi.org/10.1103/PhysRevB.99.140404}
	{\bibfield  {journal} {\bibinfo  {journal} {Phys. Rev. B}\ }\textbf {\bibinfo
			{volume} {99}},\ \bibinfo {pages} {140404} (\bibinfo {year}
		{2019})}\BibitemShut {NoStop}%
	\bibitem [{\citenamefont {Villanova}\ \emph {et~al.}(2023)\citenamefont
		{Villanova}, \citenamefont {Scheie}, \citenamefont {Tennant}, \citenamefont
		{Okamoto},\ and\ \citenamefont {Berlijn}}]{Villanova2023}%
	\BibitemOpen
	\bibfield  {author} {\bibinfo {author} {\bibfnamefont {J.~W.}\ \bibnamefont
			{Villanova}}, \bibinfo {author} {\bibfnamefont {A.~O.}\ \bibnamefont
			{Scheie}}, \bibinfo {author} {\bibfnamefont {D.~A.}\ \bibnamefont {Tennant}},
		\bibinfo {author} {\bibfnamefont {S.}~\bibnamefont {Okamoto}},\ and\ \bibinfo
		{author} {\bibfnamefont {T.}~\bibnamefont {Berlijn}},\ }\bibfield  {title}
	{\bibinfo {title} {First-principles derivation of magnetic interactions in
			the triangular quantum spin liquid candidates {$\mathrm{KYb}C{h}_{2}
				(Ch=\text{S},\mathrm{Se},\mathrm{Te})$ and $A{\mathrm{YbSe}}_{2}
				(A=\text{Na},\mathrm{Rb})$}},\ }\href
	{https://doi.org/10.1103/PhysRevResearch.5.033050} {\bibfield  {journal}
		{\bibinfo  {journal} {Phys. Rev. Res.}\ }\textbf {\bibinfo {volume} {5}},\
		\bibinfo {pages} {033050} (\bibinfo {year} {2023})}\BibitemShut {NoStop}%
	\bibitem [{\citenamefont {Paddison}(2020)}]{Paddison_2020}%
	\BibitemOpen
	\bibfield  {author} {\bibinfo {author} {\bibfnamefont {J.~A.~M.}\
			\bibnamefont {Paddison}},\ }\bibfield  {title} {\bibinfo {title} {Scattering
			signatures of bond-dependent magnetic interactions},\ }\href
	{https://doi.org/10.1103/PhysRevLett.125.247202} {\bibfield  {journal}
		{\bibinfo  {journal} {Phys. Rev. Lett.}\ }\textbf {\bibinfo {volume} {125}},\
		\bibinfo {pages} {247202} (\bibinfo {year} {2020})}\BibitemShut {NoStop}%
	\bibitem [{\citenamefont {Zhang}\ \emph {et~al.}(2021)\citenamefont {Zhang},
		\citenamefont {Li}, \citenamefont {Liu}, \citenamefont {Zhang}, \citenamefont
		{Ji}, \citenamefont {Jin}, \citenamefont {Chen}, \citenamefont {Wang},
		\citenamefont {Wang}, \citenamefont {Ma},\ and\ \citenamefont
		{Zhang}}]{Zhang_2021_Effective}%
	\BibitemOpen
	\bibfield  {author} {\bibinfo {author} {\bibfnamefont {Z.}~\bibnamefont
			{Zhang}}, \bibinfo {author} {\bibfnamefont {J.}~\bibnamefont {Li}}, \bibinfo
		{author} {\bibfnamefont {W.}~\bibnamefont {Liu}}, \bibinfo {author}
		{\bibfnamefont {Z.}~\bibnamefont {Zhang}}, \bibinfo {author} {\bibfnamefont
			{J.}~\bibnamefont {Ji}}, \bibinfo {author} {\bibfnamefont {F.}~\bibnamefont
			{Jin}}, \bibinfo {author} {\bibfnamefont {R.}~\bibnamefont {Chen}}, \bibinfo
		{author} {\bibfnamefont {J.}~\bibnamefont {Wang}}, \bibinfo {author}
		{\bibfnamefont {X.}~\bibnamefont {Wang}}, \bibinfo {author} {\bibfnamefont
			{J.}~\bibnamefont {Ma}},\ and\ \bibinfo {author} {\bibfnamefont
			{Q.}~\bibnamefont {Zhang}},\ }\bibfield  {title} {\bibinfo {title} {Effective
			magnetic hamiltonian at finite temperatures for rare-earth chalcogenides},\
	}\href {https://doi.org/10.1103/PhysRevB.103.184419} {\bibfield  {journal}
		{\bibinfo  {journal} {Phys. Rev. B}\ }\textbf {\bibinfo {volume} {103}},\
		\bibinfo {pages} {184419} (\bibinfo {year} {2021})}\BibitemShut {NoStop}%
	\bibitem [{\citenamefont {Brown}(1998)}]{BrownFF}%
	\BibitemOpen
	\bibfield  {author} {\bibinfo {author} {\bibfnamefont {P.~J.}\ \bibnamefont
			{Brown}},\ }\href {https://www.ill.eu/sites/ccsl/ffacts/} {\bibinfo {title}
		{Magnetic form factors}},\ \bibinfo {howpublished} {The Cambridge
		Crystallographic Subroutine Library} (\bibinfo {year} {1998})\BibitemShut
	{NoStop}%
	\bibitem [{\citenamefont {Robitaille}\ \emph {et~al.}(2013)\citenamefont
		{Robitaille}, \citenamefont {Tollerud}, \citenamefont {Greenfield},
		\citenamefont {Droettboom}, \citenamefont {Bray}, \citenamefont {Aldcroft},
		\citenamefont {Davis}, \citenamefont {Ginsburg}, \citenamefont
		{Price-Whelan}, \citenamefont {Kerzendorf} \emph
		{et~al.}}]{robitaille2013astropy}%
	\BibitemOpen
	\bibfield  {author} {\bibinfo {author} {\bibfnamefont {T.~P.}\ \bibnamefont
			{Robitaille}}, \bibinfo {author} {\bibfnamefont {E.~J.}\ \bibnamefont
			{Tollerud}}, \bibinfo {author} {\bibfnamefont {P.}~\bibnamefont
			{Greenfield}}, \bibinfo {author} {\bibfnamefont {M.}~\bibnamefont
			{Droettboom}}, \bibinfo {author} {\bibfnamefont {E.}~\bibnamefont {Bray}},
		\bibinfo {author} {\bibfnamefont {T.}~\bibnamefont {Aldcroft}}, \bibinfo
		{author} {\bibfnamefont {M.}~\bibnamefont {Davis}}, \bibinfo {author}
		{\bibfnamefont {A.}~\bibnamefont {Ginsburg}}, \bibinfo {author}
		{\bibfnamefont {A.~M.}\ \bibnamefont {Price-Whelan}}, \bibinfo {author}
		{\bibfnamefont {W.~E.}\ \bibnamefont {Kerzendorf}}, \emph {et~al.},\
	}\bibfield  {title} {\bibinfo {title} {Astropy: A community python package
			for astronomy},\ }\href {https://doi.org/10.1051/0004-6361/201322068}
	{\bibfield  {journal} {\bibinfo  {journal} {Astronomy \& Astrophysics}\
		}\textbf {\bibinfo {volume} {558}},\ \bibinfo {pages} {A33} (\bibinfo {year}
		{2013})}\BibitemShut {NoStop}%
	\bibitem [{\citenamefont {Sullivan}\ and\ \citenamefont
		{Seidel}(1968)}]{Sullivan_1968}%
	\BibitemOpen
	\bibfield  {author} {\bibinfo {author} {\bibfnamefont {P.~F.}\ \bibnamefont
			{Sullivan}}\ and\ \bibinfo {author} {\bibfnamefont {G.}~\bibnamefont
			{Seidel}},\ }\bibfield  {title} {\bibinfo {title} {Steady-state,
			ac-temperature calorimetry},\ }\href
	{https://doi.org/10.1103/PhysRev.173.679} {\bibfield  {journal} {\bibinfo
			{journal} {Phys. Rev.}\ }\textbf {\bibinfo {volume} {173}},\ \bibinfo {pages}
		{679} (\bibinfo {year} {1968})}\BibitemShut {NoStop}%
	\bibitem [{\citenamefont {Syromyatnikov}(2023)}]{syromyatnikov2022unusual}%
	\BibitemOpen
	\bibfield  {author} {\bibinfo {author} {\bibfnamefont {A.}~\bibnamefont
			{Syromyatnikov}},\ }\bibfield  {title} {\bibinfo {title} {Unusual dynamics of
			spin-12 antiferromagnets on the triangular lattice in magnetic field},\
	}\href {https://doi.org/https://doi.org/10.1016/j.aop.2023.169342} {\bibfield
		{journal} {\bibinfo  {journal} {Annals of Physics}\ }\textbf {\bibinfo
			{volume} {454}},\ \bibinfo {pages} {169342} (\bibinfo {year}
		{2023})}\BibitemShut {NoStop}%
	\bibitem [{\citenamefont {Kresse}\ and\ \citenamefont
		{Furthm\"uller}(1996{\natexlab{a}})}]{VASP1}%
	\BibitemOpen
	\bibfield  {author} {\bibinfo {author} {\bibfnamefont {G.}~\bibnamefont
			{Kresse}}\ and\ \bibinfo {author} {\bibfnamefont {J.}~\bibnamefont
			{Furthm\"uller}},\ }\bibfield  {title} {\bibinfo {title} {Efficient iterative
			schemes for ab initio total-energy calculations using a plane-wave basis
			set},\ }\href {https://doi.org/10.1103/PhysRevB.54.11169} {\bibfield
		{journal} {\bibinfo  {journal} {Phys. Rev. B}\ }\textbf {\bibinfo {volume}
			{54}},\ \bibinfo {pages} {11169} (\bibinfo {year}
		{1996}{\natexlab{a}})}\BibitemShut {NoStop}%
	\bibitem [{\citenamefont {Kresse}\ and\ \citenamefont
		{Furthm\"uller}(1996{\natexlab{b}})}]{VASP2}%
	\BibitemOpen
	\bibfield  {author} {\bibinfo {author} {\bibfnamefont {G.}~\bibnamefont
			{Kresse}}\ and\ \bibinfo {author} {\bibfnamefont {J.}~\bibnamefont
			{Furthm\"uller}},\ }\bibfield  {title} {\bibinfo {title} {Efficiency of
			ab-initio total energy calculations for metals and semiconductors using a
			plane-wave basis set},\ }\href
	{https://doi.org/https://doi.org/10.1016/0927-0256(96)00008-0} {\bibfield
		{journal} {\bibinfo  {journal} {Comput. Mater. Sci.}\ }\textbf {\bibinfo
			{volume} {6}},\ \bibinfo {pages} {15} (\bibinfo {year}
		{1996}{\natexlab{b}})}\BibitemShut {NoStop}%
	\bibitem [{\citenamefont {Perdew}\ \emph {et~al.}(1996)\citenamefont {Perdew},
		\citenamefont {Burke},\ and\ \citenamefont {Ernzerhof}}]{GGA}%
	\BibitemOpen
	\bibfield  {author} {\bibinfo {author} {\bibfnamefont {J.~P.}\ \bibnamefont
			{Perdew}}, \bibinfo {author} {\bibfnamefont {K.}~\bibnamefont {Burke}},\ and\
		\bibinfo {author} {\bibfnamefont {M.}~\bibnamefont {Ernzerhof}},\ }\bibfield
	{title} {\bibinfo {title} {{Generalized Gradient Approximation Made
				Simple}},\ }\href {https://doi.org/10.1103/PhysRevLett.77.3865} {\bibfield
		{journal} {\bibinfo  {journal} {Phys. Rev. Lett.}\ }\textbf {\bibinfo
			{volume} {77}},\ \bibinfo {pages} {3865} (\bibinfo {year}
		{1996})}\BibitemShut {NoStop}%
	\bibitem [{\citenamefont {Bl\"ochl}(1994)}]{PAW1}%
	\BibitemOpen
	\bibfield  {author} {\bibinfo {author} {\bibfnamefont {P.~E.}\ \bibnamefont
			{Bl\"ochl}},\ }\bibfield  {title} {\bibinfo {title} {Projector augmented-wave
			method},\ }\href {https://doi.org/10.1103/PhysRevB.50.17953} {\bibfield
		{journal} {\bibinfo  {journal} {Phys. Rev. B}\ }\textbf {\bibinfo {volume}
			{50}},\ \bibinfo {pages} {17953} (\bibinfo {year} {1994})}\BibitemShut
	{NoStop}%
	\bibitem [{\citenamefont {Kresse}\ and\ \citenamefont {Joubert}(1999)}]{PAW2}%
	\BibitemOpen
	\bibfield  {author} {\bibinfo {author} {\bibfnamefont {G.}~\bibnamefont
			{Kresse}}\ and\ \bibinfo {author} {\bibfnamefont {D.}~\bibnamefont
			{Joubert}},\ }\bibfield  {title} {\bibinfo {title} {From ultrasoft
			pseudopotentials to the projector augmented-wave method},\ }\href
	{https://doi.org/10.1103/PhysRevB.59.1758} {\bibfield  {journal} {\bibinfo
			{journal} {Phys. Rev. B}\ }\textbf {\bibinfo {volume} {59}},\ \bibinfo
		{pages} {1758} (\bibinfo {year} {1999})}\BibitemShut {NoStop}%
	\bibitem [{\citenamefont {Mostofi}\ \emph {et~al.}(2014)\citenamefont
		{Mostofi}, \citenamefont {Yates}, \citenamefont {Pizzi}, \citenamefont {Lee},
		\citenamefont {Souza}, \citenamefont {Vanderbilt},\ and\ \citenamefont
		{Marzari}}]{W90}%
	\BibitemOpen
	\bibfield  {author} {\bibinfo {author} {\bibfnamefont {A.~A.}\ \bibnamefont
			{Mostofi}}, \bibinfo {author} {\bibfnamefont {J.~R.}\ \bibnamefont {Yates}},
		\bibinfo {author} {\bibfnamefont {G.}~\bibnamefont {Pizzi}}, \bibinfo
		{author} {\bibfnamefont {Y.-S.}\ \bibnamefont {Lee}}, \bibinfo {author}
		{\bibfnamefont {I.}~\bibnamefont {Souza}}, \bibinfo {author} {\bibfnamefont
			{D.}~\bibnamefont {Vanderbilt}},\ and\ \bibinfo {author} {\bibfnamefont
			{N.}~\bibnamefont {Marzari}},\ }\bibfield  {title} {\bibinfo {title} {An
			updated version of wannier90: A tool for obtaining maximally-localised
			wannier functions},\ }\href
	{https://doi.org/https://doi.org/10.1016/j.cpc.2014.05.003} {\bibfield
		{journal} {\bibinfo  {journal} {Comput. Phys. Commun.}\ }\textbf {\bibinfo
			{volume} {185}},\ \bibinfo {pages} {2309 } (\bibinfo {year}
		{2014})}\BibitemShut {NoStop}%
	\bibitem [{\citenamefont {Marzari}\ and\ \citenamefont
		{Vanderbilt}(1997)}]{Marzari1997}%
	\BibitemOpen
	\bibfield  {author} {\bibinfo {author} {\bibfnamefont {N.}~\bibnamefont
			{Marzari}}\ and\ \bibinfo {author} {\bibfnamefont {D.}~\bibnamefont
			{Vanderbilt}},\ }\bibfield  {title} {\bibinfo {title} {{Maximally localized
				generalized Wannier functions for composite energy bands}},\ }\href
	{https://doi.org/10.1103/PhysRevB.56.12847} {\bibfield  {journal} {\bibinfo
			{journal} {Phys. Rev. B}\ }\textbf {\bibinfo {volume} {56}},\ \bibinfo
		{pages} {12847} (\bibinfo {year} {1997})}\BibitemShut {NoStop}%
	\bibitem [{\citenamefont {Souza}\ \emph {et~al.}(2001)\citenamefont {Souza},
		\citenamefont {Marzari},\ and\ \citenamefont {Vanderbilt}}]{Souza2001}%
	\BibitemOpen
	\bibfield  {author} {\bibinfo {author} {\bibfnamefont {I.}~\bibnamefont
			{Souza}}, \bibinfo {author} {\bibfnamefont {N.}~\bibnamefont {Marzari}},\
		and\ \bibinfo {author} {\bibfnamefont {D.}~\bibnamefont {Vanderbilt}},\
	}\bibfield  {title} {\bibinfo {title} {{Maximally localized Wannier functions
				for entangled energy bands}},\ }\href
	{https://doi.org/10.1103/PhysRevB.65.035109} {\bibfield  {journal} {\bibinfo
			{journal} {Phys. Rev. B}\ }\textbf {\bibinfo {volume} {65}},\ \bibinfo
		{pages} {035109} (\bibinfo {year} {2001})}\BibitemShut {NoStop}%
\end{thebibliography}

%





\end{document}